\documentclass[12pt]{article}
\usepackage[letterpaper, margin=1in]{geometry}
\usepackage{setspace}
\usepackage{natbib}
\usepackage{amsmath}
\usepackage{graphicx}
\usepackage{hyperref}
\usepackage{caption}
\usepackage{subcaption}
\usepackage{float}
\usepackage{subfiles}
\usepackage{makecell}
\usepackage{xcolor}
\usepackage[compact]{titlesec}
\titlespacing{\section}{0pt}{-1.5ex}{-0ex}
\titlespacing{\subsection}{0pt}{-1.5ex}{0ex}
\titlespacing{\subsubsection}{0pt}{-0ex}{0ex}
\providecommand{\keywords}[1]{\textbf{Key Words:} #1}
\newtheorem{proposition}{Proposition}

\date{\vspace{-5ex}}

\usepackage{authblk}
\author[1]{Qing Yin}
\author[1]{Jong-Hyeon Jeong}
\author[2]{Xu Qin}
\author[3]{Shyamal D Peddada*}
\author[4]{Jennifer J Adibi*}

\affil[1]{Department of Biostatistics, University of Pittsburgh}
\affil[2]{Department of Health and Human Development, University of Pittsburgh}
\affil[3]{Biostatistics and Bioinformatics Branch, NICHD}
\affil[4]{Department of Epidemiology, University of Pittsburgh}

\title{Mediation Analysis using Semi-parametric Shape-Restricted Regression with Applications}

\begin{document}
\maketitle

\bibliographystyle{abbrvnat}
\setcitestyle{authoryear,open={(},close={)}}

\abstract{Often linear regression is used to perform mediation analysis. However, in many instances, the underlying relationships may not be linear, as in the case of placental-fetal hormones and fetal development. Although, the exact functional form of the relationship may be unknown, one may hypothesize the general shape of the relationship. For these reasons, we develop a novel shape-restricted inference-based methodology for conducting mediation analysis. This work is motivated by an application in fetal endocrinology where researchers are interested in understanding the effects of pesticide application on birth weight, with human chorionic gonadotropin (hCG) as the mediator. We assume a practically plausible set of nonlinear effects of hCG on the birth weight and a linear relationship between pesticide exposure and hCG, with both exposure-outcome and exposure-mediator models being linear in the confounding factors. Using the proposed methodology on a population-level prenatal screening program data, with hCG as the mediator, we discovered that, while the natural direct effects suggest a positive association between pesticide application and birth weight, the natural indirect effects were negative.}


\section{Introduction}
\label{sec:1}
Population level life course data, originating during the fetal period, were originally used to develop the developmental origins of health and disease hypothesis \citep{barker2003}. For example, infants whose mothers were exposed to famine in Europe in the last two trimesters during World War II tended to have lower birth weight and this was consequential for their subsequent risk of schizophrenia and adult cardiovascular disease \citep{hoek1998, vanabeelen2011}. Later in his career, David Barker actively pursued measures of the placenta as a mediator to support his theory \citep{barker2013}. However, these associations are challenging to interpret causally due to the many potential explanations and the lack of a causal mediator that can conceptually and quantitatively link the exposure and the outcome. Recently, \cite{adibi2021b} established a methodology accounting for the role of placenta in a causal framework to address this gap.

Even though the analysis of a direct exposure-outcome effect is usually intuitive and easy to interpret, inclusion of a putative mediator can further strengthen causal inference and offer key insights to the reporting of an association. In some physiologic examples, such as pregnancy, it is implausible to not consider the role of the mediator. The placenta, a pregnancy-specific organ, is a physical and biologic interface between the mother and the fetus, and plays an important role in fetal growth \citep{benirschke2000}. For example, the placenta provides the fetus with nutrients. It can serve as a barrier and protect the fetus from some environmental toxins \citep{benirschke2000}. 

The placenta supports development of the fetus by way of producing growth factors and hormones. We hypothesize that environmental factors can change the quantities, proportions or timing of the placental hormones. In turn, these changes can alter development of organ systems. In the case of sex steroid hormones, the effects of the hormones are mediated by nuclear receptors and generally follow a U-shaped pattern \citep{Li2007}. In the case of gonadotropins, the effects are mediated by membrane receptors and can cause distinct effects at low doses that differ from those at high doses, as has been studied with gonadotropin releasing hormone. See for example, \cite{Vandenberg2013} and references therein.

The aim of this paper is to address two problems in developmental origins of health and disease (DOHaD) epidemiology: (1) the need for a causal mediation framework, and (2) the need for a method that accommodates nonlinearity in the mediator-outcome relationship. Regression-based mediation analysis has been developed to address this problem in epidemiology \citep{vanderweele2015, baron1986}. However, this strategy assumes a linear relationship between the mediator and the outcome. In such cases as described above, linear models may potentially result in model misspecification, biased estimation, and hence incorrect conclusions. 

\begin{figure}
\centering
\includegraphics[width=0.6\textwidth]{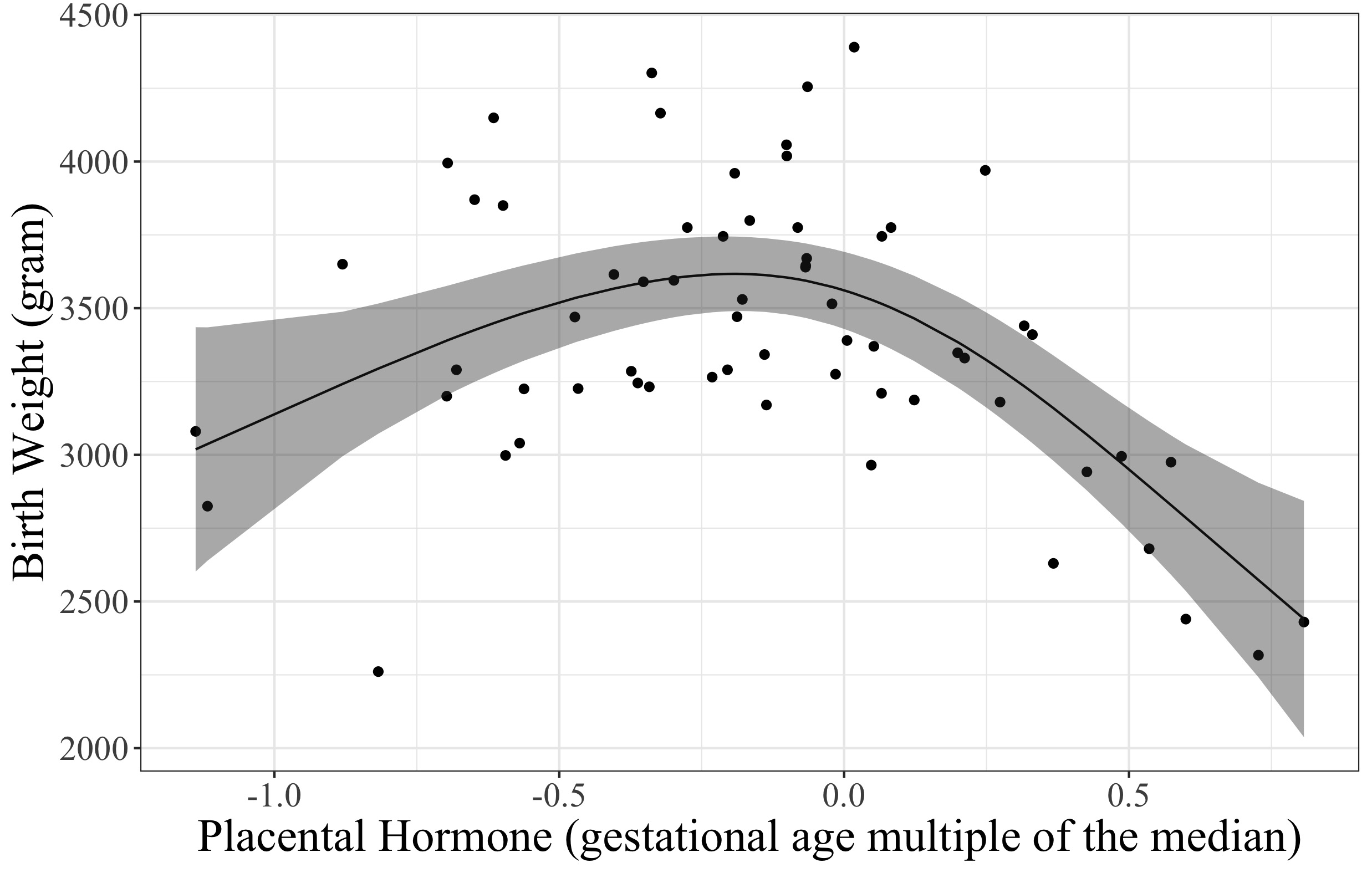}
\caption{The relationship between a placental hormone (expressed as a gestational age multiple of median) and infant birth weight.}
\label{fig:1}
\end{figure}

The application of the generalized additive model (GAM) was proposed by \cite{imai2010} to address the nonlinearity, where the direct and indirect effects are estimated using simulations. Although the GAM provides flexibility in modeling nonlinear relationships, the main effect is usually fitted using a smoothing curve, which is hard to be parametrized, and in many instances, the patterns of mean response may not be arbitrary. A simulation-based method was used for mediation analysis \citep{tingley2014}, but the procedure is computationally intensive and does not allow for conditioning on confounding variables (thus, the controlled direct effect (CDE) may not be calculated). May and Bigelow \citep{May2005} outlined methodological challenges and caveats when statistically modeling nonlinear associations using standard approaches such as splines. Hence, we aim to develop a flexible semi-parametric shape-restricted inference-based methodology to overcome some of the modelling challenges. Unless one is interested in modelling dynamic systems such as circadian clock or cell-cycle, in many applications, it may be sufficient to model the relationships assuming monotonic, convex or concave shapes. An example of such nonlinear relationship is shown in Figure \ref{fig:1}, where the relationship between the placental hormone and infant birth weight is concave. Hence in this paper we limit our focus on these classes of models. Our approach makes a compromise between the standard linear or polynomial regressions, which are rigid in shapes, and GAMs, which are too flexible by allowing arbitrary shapes and computationally intensive.

The paper is organized as follows. Using semi-parametric shape-restricted regression splines \citep{meyer2008, meyer2018, yin2021}, we develop outcome and mediator models in Section \ref{sec:2}. The estimated parameters from these models are then used to estimate the direct and indirect effects of various factors and their confidence intervals. The performance of these estimators are evaluated using simulation studies in Section \ref{sec:3}. In Section \ref{sec:4}, the proposed methodology is applied to a population-level prenatal screening program data set to describe the role of placental hormones as mediators in the relationship between pesticide application and the birth weight of infants. Concluding remarks are provided in Section \ref{sec:5}. All mathematical details are provided in the supplementary text.
\section{Methodology}
\label{sec:2}

\subsection{Background and Notations}
\label{sec:2.1}
Throughout this paper, the exposure variable for an individual will be denoted by $A$, $M$ denotes the mediator, $C$ denotes the confounder, and $Y$ denotes the outcome variable. For a given individual with exposure $A = a$, the outcome is denoted by $Y_a$ and the mediator is denoted by $M_a$, and $Y_{am}$ denotes the outcome of an individual whose exposure $A = a$ and mediator $M = m$. The controlled direct effect (CDE), the natural direct effect (NDE) and the natural indirect effect (NIE) are denoted by $Y_{am}-Y_{a^*m}$, $Y_{aM_{a^*}}-Y_{a^*M_{a^*}}$ and $Y_{aM_a}-Y_{aM_{a^*}}$, respectively. For the identifiability of the CDE, we assume that there is no unmeasured exposure-outcome confounding, i.e., $Y_{am} \perp (independent)  A | C$, and there is no unmeasured mediator-outcome confounding, i.e., $Y_{am} \perp  M | {A, C}$. For the identifiability of the NDE and NIE, besides the two assumptions above, we additionally assume that there is no unmeasured exposure-mediator confounding, i.e., $M_a \perp A | C$, and there is no mediator-outcome confounding affected by exposure, i.e., $Y_{am} \perp M_{a^*} | C$.

For a real number $x$, and a sequence of knots $t = \{t_1, t_2, \ldots, t_{n+k}\}$, $M_i(x|k,t)$ denotes the $k^{th}$ order M-spline basis function, which is a piece-wise polynomial of degree $k - 1$ \citep{curry1966}. The corresponding I-spline \citep{ramsay1988} and C-spline \citep{meyer2018} basis functions are given by $I_i(x|k,t) = \int_L^x M_i(u|k,t)du$ and $C_i(x|k,t) = \int_L^x I_i(u|k,t)du$, respectively.

The methodology developed in this paper relies on the general framework introduced in \cite{meyer2018} and \cite{yin2021}, which uses quadratic I-splines and cubic C-splines (see supplementary text for details). The quadratic I-splines are of interest in shape-restricted regression because a linear combination of quadratic I-spline basis functions is non-decreasing if and only if the coefficients are non-negative. The regression function using I-splines is estimated by a linear combination of the basis functions, the constant function and other covariates \citep{meyer2018}. The cubic C-splines are of interest in shape-restricted regression because a linear combination of cubic C-spline basis functions is convex if and only if the coefficients are non-negative. The regression function using C-splines is estimated by a linear combination of the basis functions, the constant function, the identity function and other covariates \citep{meyer2018}.

\subsection{Model}
\label{sec:2.2}
Suppose that an interaction between exposure and mediator exists, and the relationships between mediator and outcome in both exposure and non-exposure groups are potentially nonlinear. Then the exposure-outcome relationship is modeled by 
\begin{equation} \label{eq:2.2:1}
    Y = \beta_0 + \beta_1 A + f_1(M) A + f_2(M)(1 - A) + \beta_4 C + \epsilon_1,
\end{equation}
where $f_1(M)$ is the function of the mediator for the exposure group, $f_2(M)$ is the function of the mediator for the non-exposure group, and $\epsilon_1 \sim N(0, \sigma_1^2)$, and the exposure-mediator model will be 
\begin{equation} \label{eq:2.2:2}
    M = \gamma_0 + \gamma_1 A + \gamma_2 C + \epsilon_2,
\end{equation}
where $\epsilon_2 \sim N(0, \sigma_2^2)$.  

In the case of quadratic I-splines, $f_1(M) = \sum_{i=1}^k \beta_{2i} I_i(M|2, t)$ and $f_2(M) = \sum_{i=1}^k$ $\beta_{3i} I_i(M|2, t)$, and in case of cubic C-splines, $f_1(M) = \beta_{20}M + \sum_{i=1}^k \beta_{2i} C_i(M|2, t)$ and $f_2(M) = \beta_{30}M + \sum_{i=1}^k \beta_{3i} C_i(M|2, t)$, reflecting the structures of the I- and C-spline basis functions in the supplementary text. Define $IS = [I_1(M|2, t), ..., I_k(M|2, t)]$ and $CS = [M, C_1(M|2, t), ..., C_k(M|2, t)]$. Denote the symbol ``$\bullet$'' for the face-splitting product (for $A$ = $\begin{bmatrix} a_1 \\ a_2 \end{bmatrix}$ and $B$ = $\begin{bmatrix} b_{11} & b_{12} \\ b_{21} & b_{22} \end{bmatrix}$, the face-splitting product operates as $A \bullet B$ = $\begin{bmatrix} a_1 b_{11} & a_1 b_{12} \\ a_2 b_{21} & a_2 b_{22} \end{bmatrix}$), and let $\beta_2 = [\beta_{21}, ..., \beta_{2k}]$ if the corresponding matrix is $IS \bullet A$, or $\beta_2 = [\beta_{20}, \beta_{21}, ..., \beta_{2k}]$ if the corresponding matrix is $CS \bullet A$, and $\beta_3 = [\beta_{31}, ..., \beta_{3k}]$ if the corresponding matrix is $IS \bullet (1 - A)$, or $\beta_3 = [\beta_{30}, \beta_{31}, ..., \beta_{3k}]$ if the corresponding matrix is $CS \bullet (1 - A)$.

If $f_1(M)$ and $f_2(M)$ are both monotonic, then they are fitted using the I-splines, and model (\ref{eq:2.2:1}) is given by
\begin{equation} \label{eq:2.2:3}
\begin{aligned}
    Y &= [1, A, IS \bullet A, IS \bullet (1 - A), C] [\beta_0, \beta_1, \beta_{2}, \beta_{3}, \beta_4]^T + \epsilon_1.
\end{aligned}
\end{equation}
If $f_1(M)$ and $f_2(M)$ are both convex (or concave), then they are fitted using the C-splines, and the model (\ref{eq:2.2:1}) is given by
\begin{equation} \label{eq:2.2:4}
\begin{aligned}
    Y &= [1, A, CS \bullet A, CS \bullet (1 - A), C] [\beta_0, \beta_1, \beta_{2}, \beta_{3}, \beta_4]^T + \epsilon_1.
\end{aligned}
\end{equation}
If $f_1(M)$ is monotonic but $f_2(M)$ is convex (or concave), then $f_1(M)$ is fitted using I-splines and $f_2(M)$ is fitted using C-splines, and the model (\ref{eq:2.2:1}) is given by
\begin{equation} \label{eq:2.2:5}
\begin{aligned}
    Y &= [1, A, IS \bullet A, CS \bullet (1 - A), C] [\beta_0, \beta_1, \beta_{2}, \beta_{3}, \beta_4]^T + \epsilon_1.
\end{aligned}
\end{equation}
If $f_1(M)$ is convex (or concave) and $f_2(M)$ is monotonic, then $f_1(M)$ is fitted using C-splines and $f_2(M)$ is fitted using I-splines, and the model (\ref{eq:2.2:1}) is given by
\begin{equation} \label{eq:2.2:6}
\begin{aligned}
    Y &= [1, A, CS \bullet A, IS \bullet (1 - A), C] [\beta_0, \beta_1, \beta_{2}, \beta_{3}, \beta_4]^T + \epsilon_1.
\end{aligned}
\end{equation}

\subsection{Statistical Inference}
\label{sec:2.3}
Regression parameters of the model (\ref{eq:2.2:1}) are estimated along the lines of \cite{meyer2018}, but the method is modified to account for factor-by-curve interaction. 

Define $CS = [CS_1, CS_2]$, where $CS_1 = M$ and $CS_2 = [C_1(M|2, t), ..., C_k(M|2, t)]$. For model (\ref{eq:2.2:3}), $W_0 = [1], W = [A, C], Z_1 = [IS \bullet A], Z_0 = [IS \bullet (1 - A)] \text{ and } Z = [Z_1, Z_0]$; for model (\ref{eq:2.2:4}), $W_0 = [1, CS_1 \bullet A, CS_1 \bullet (1 - A)], W = [A, C], Z_1 = [CS_2 \bullet A], Z_0 = [CS_2 \bullet(1 - A)] \text{ and } Z = [Z_1, Z_0]$; for model (\ref{eq:2.2:5}), $W_0 = [1, CS_1 \bullet (1 - A)], W = [A, C], Z_1 = [IS \bullet A], Z_0 = [CS_2 \bullet (1 - A)] \text{ and } Z = [Z_1, Z_0]$; for model (\ref{eq:2.2:6}), $W_0 = [1, CS_1 \bullet A], W = [A, C], Z_1 = [CS_2 \bullet A], Z_0 = [IS \bullet (1 - A)] \text{ and } Z = [Z_1, Z_0]$.

Let $V = [W_0, W]$, $P_V = V (V^T V)^{-1} V^T$, the orthogonal projection operator onto the column space of $V^T$, and $\Delta = (I - P_V)Z$. Using the hinge algorithm for cone projection \citep{meyer2013}, a subset of columns of $\Delta$ are determined. We then keep the corresponding columns of $Z$ and estimate the parameters of the model (\ref{eq:2.2:1}) using the ordinary least squares. The parameters corresponding to the eliminated columns of $Z$ are estimated as 0. During the process of the hinge algorithm, if the signs of coefficients for the exposure or non-exposure group splines are assumed to be non-positive, i.e., the curve is assumed to be decreasing or concave, then we will use $-IS$ or $-CS_2$ instead of $IS$ or $CS_2$. We estimate parameters of the model (\ref{eq:2.2:2}) using ordinary least squares.

Under the assumptions described in Section \ref{sec:2.1}, we parametrize the expected controlled direct effect, natural direct effect and natural indirect effect as follows in Proposition \ref{prop:2.3:1}. The detailed derivation is provided in the supplementary text. 

\begin{proposition} \label{prop:2.3:1}
Suppose (1) for an individual with actual exposure $A = a$, the actual outcome $Y$ is $Y_a$ (consistency), (2) $Y_{am} \perp  A | C$, (3) $Y_{am} \perp  M | {A, C}$, (4) $M_a \perp A | C$ and (5) $Y_{am} \perp M_{a^*} | C$, and suppose (\ref{eq:2.2:1}) and (\ref{eq:2.2:2}) are correctly specified, then the expected controlled direct effect, natural direct effect and natural indirect effect, conditional on $C = c$, are given by
\begin{equation} \label{eq:2.3:1}
    E[Y_{am} - Y_{a^*m}|c] = (\beta_1 + f_1(m) - f_2(m))(a - a^*),
\end{equation}
\begin{equation} \label{eq:2.3:2}
    E[Y_{aM_{a^*}}-Y_{a^*M_{a^*}}|c] = (\beta_1 + E[f_1(M)|a^*, c] - E[f_2(M)|a^*, c])(a - a^*),
\end{equation}
and
\begin{equation} \label{eq:2.3:3}
    E[Y_{aM_a}-Y_{aM_{a^*}}|c] = a(E[f_1(M)|a, c] - E[f_1(M)|a^*, c]) + (1 - a)(E[f_2(M)|a, c] - E[f_2(M)|a^*, c]),
\end{equation}
respectively.
\end{proposition}

According to Proposition \ref{prop:2.3:1}, the expected CDE is a function of $\beta_1$, $\beta_2$ and $\beta_3$, the expected NDE is a function of $\beta_1, \beta_2$, $\beta_3$, $\gamma_0$, $\gamma_1$, $\gamma_2$ and $\sigma_2^2$, and the expected NIE is a function of $\beta_2$, $\gamma_0$, $\gamma_1$, $\gamma_2$ and $\sigma_2^2$. Thus, these mediation effects are derived by plugging in the suitable least squares or constrained estimators. We apply the delta method to obtain the asymptotic variances of different mediation effects. All unknown parameters in the asymptotic variance expressions are replaced by the point estimates obtained from the least squares. The 95\% confidence intervals are derived using the standard formula $(g(\hat{\theta}) - z_{0.975} \sqrt{\widehat{var(g(\hat{\theta}))}}, g(\hat{\theta}) + z_{0.975} \sqrt{\widehat{var(g(\hat{\theta}))}})$. The technical details are provided in Proposition 3 in the supplementary text, and the computational details are also provided in the supplementary text.
\section{Simulation Study}
\label{sec:3}
We evaluate the performance of our methodology in terms of the coverage probability (or the probability that the 95\% confidence interval contains the true mediation effect), average absolute relative bias and average mean squared error (MSE). Data were simulated to resemble the state-wide prenatal screening program data presented in the Application section (Section \ref{sec:4}). The simulated data set contains 500 observations and 10 variables. The confounding variables are age (randomly sampled from 18 to 40 years with an increment 0.5 years), inverse maternal weight (randomly sampled from 0.0020 to 0.0143 lbs$^{-1} \times 10^3$ with an increment of 0.0001 lbs$^{-1} \times 10^3$), race (randomly sampled from race 1 to race 5 with probabilities 0.46, 0.28, 0.13, 0.1 and 0.03 respectively), season of blood draw (randomly sampled from season 1 to season 4 with the same probability 0.25), smoking status (randomly sampled from a binomial distribution with 5\% chance of smoking), ovum donor status (randomly sampled from a binomial distribution with 2\% chance of donation) and pre-existing diabetes status (randomly sampled from a binomial distribution with 5\% chance of diabetes). The exposure variable is pesticide exposure (randomly sampled from a binomial distribution with 50\% chance of being exposed). The mediator variable is hormone (gestational age multiple of median, calculated via exposure-mediator model). The outcome variable is birth weight (grams, calculated via exposure-outcome model). 

\begin{figure}
\centering
\subfloat[Pattern 1: $f_1(M) = \frac{-6(M - 5/3)^2}{5} + 100$ and $f_2(M) = \frac{50e^{6M/5}}{2 + e^{6M/5}} + 50$]{\label{fig:3:1.a}\includegraphics[width=.4\linewidth]{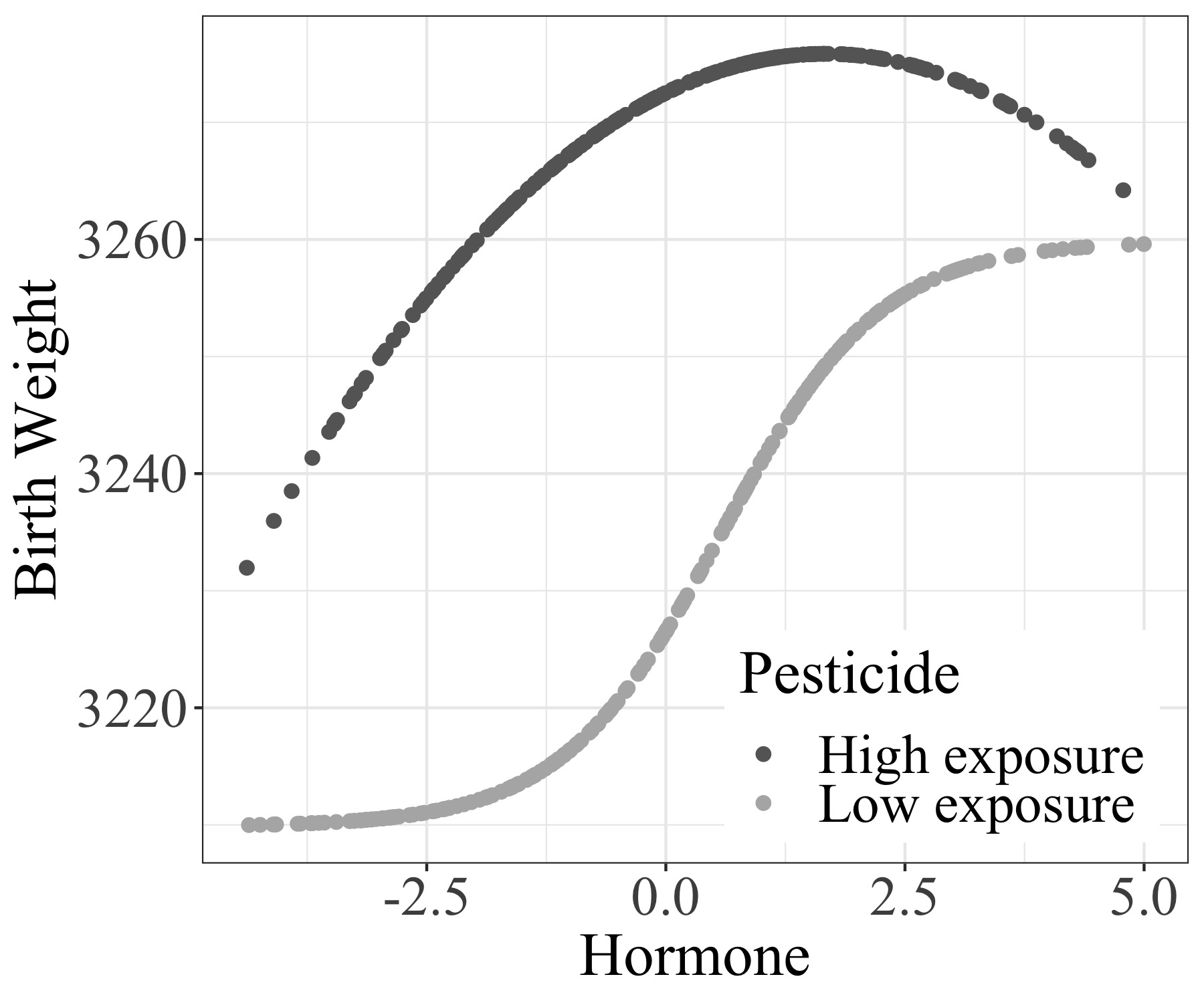}}\hspace{1.5cm}
\subfloat[Coverage Probability under Pattern 1]{\label{fig:3:1.b}\includegraphics[width=.4\linewidth]{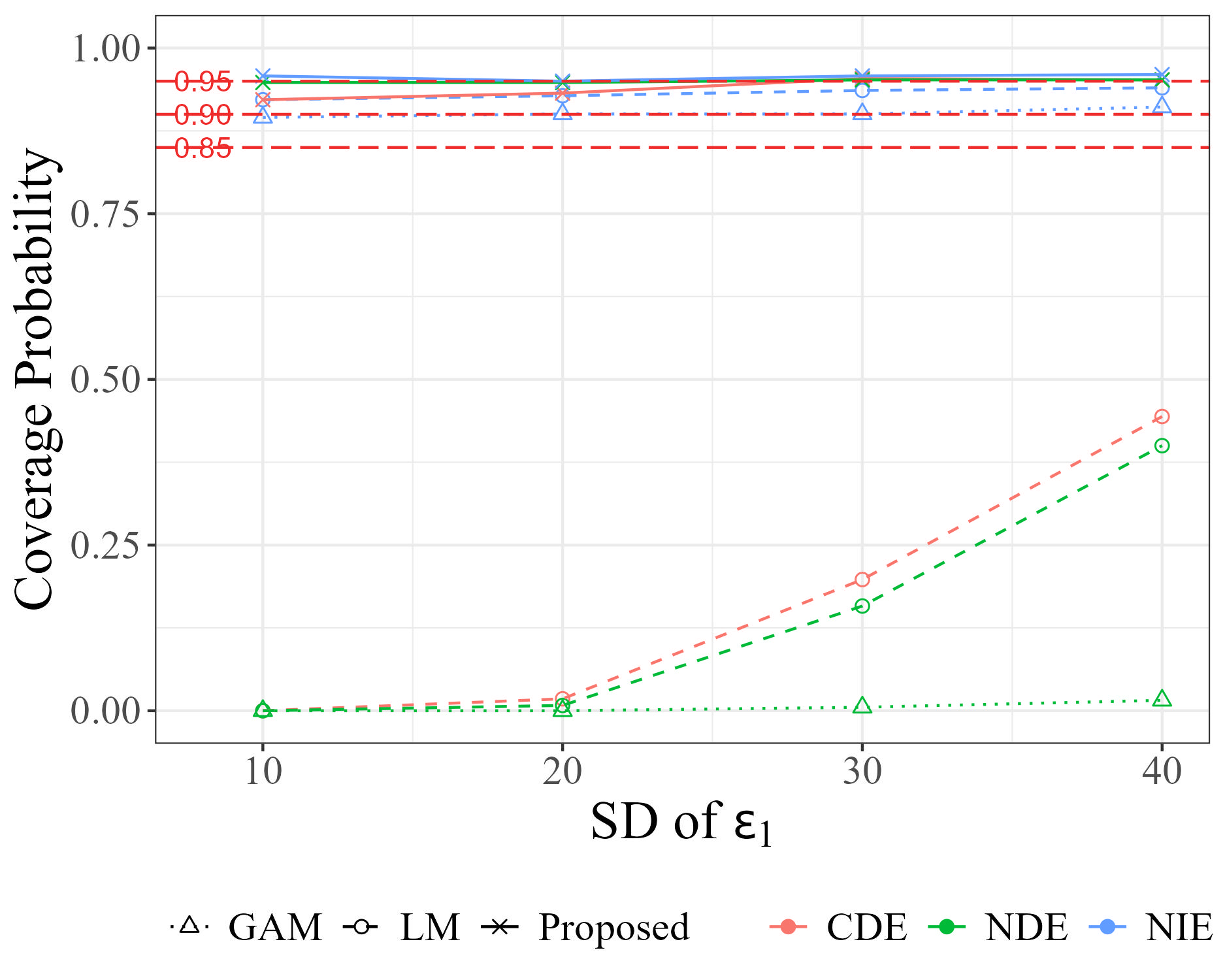}}\par
\subfloat[Pattern 2: $f_1(M) = \frac{-e^M - 100M^2}{50} + 100$ and $f_2(M) = \frac{-6(M + 5/3)^2}{5} + 100$]{\label{fig:3:2.a}\includegraphics[width=.42\linewidth]{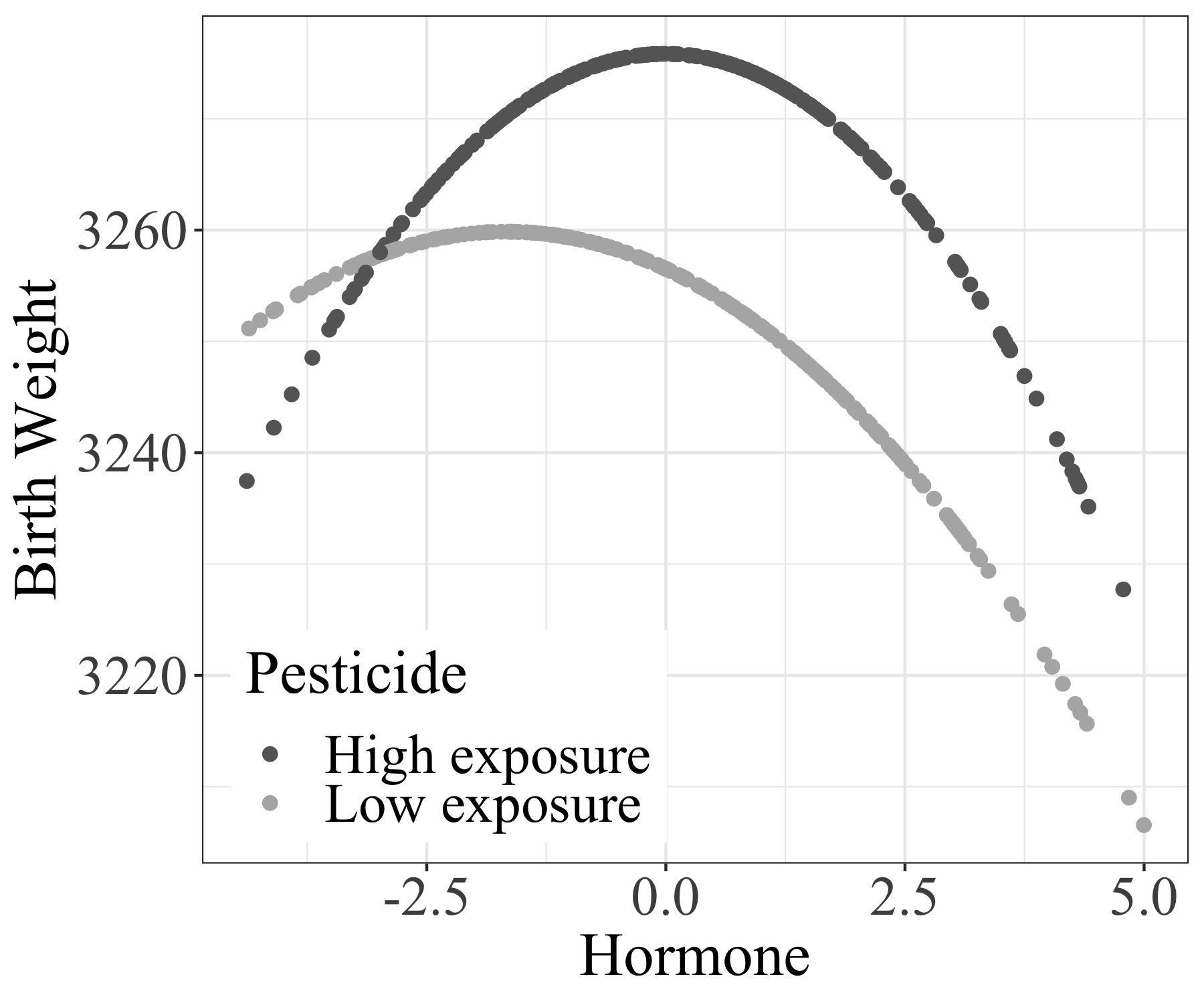}}\hspace{1.5cm}
\subfloat[Coverage Probability under pattern 2]{\label{fig:3:2.b}\includegraphics[width=.4\linewidth]{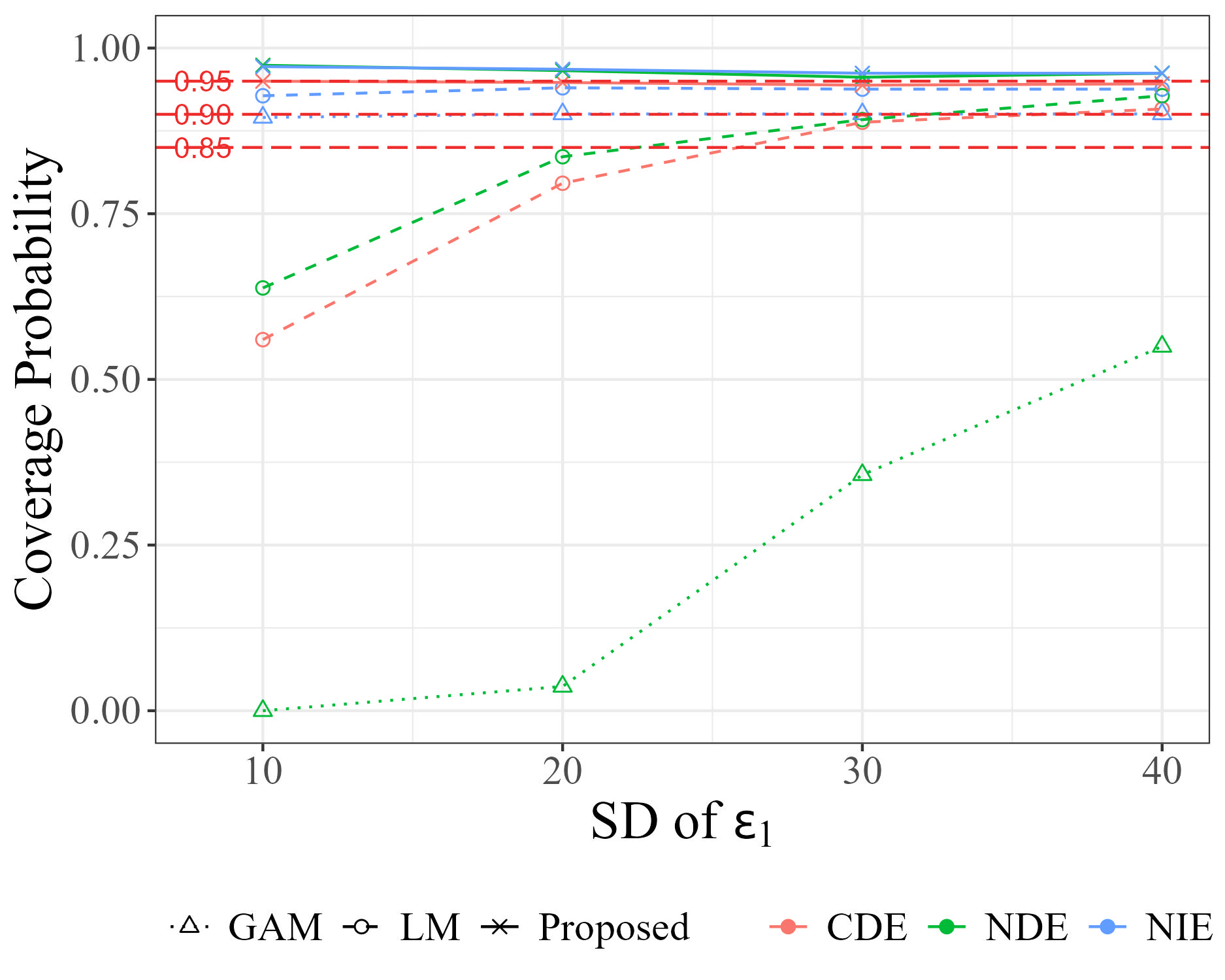}}\par
\subfloat[Pattern 3: $f_1(M) = \frac{50e^{6M/5}}{2 + e^{6M/5}} + 50$ and $f_2(M) = 300 \text{ln} (-e^{M/2} + M + 40) - 1000$]{\label{fig:3:3.a}\includegraphics[width=.4\linewidth]{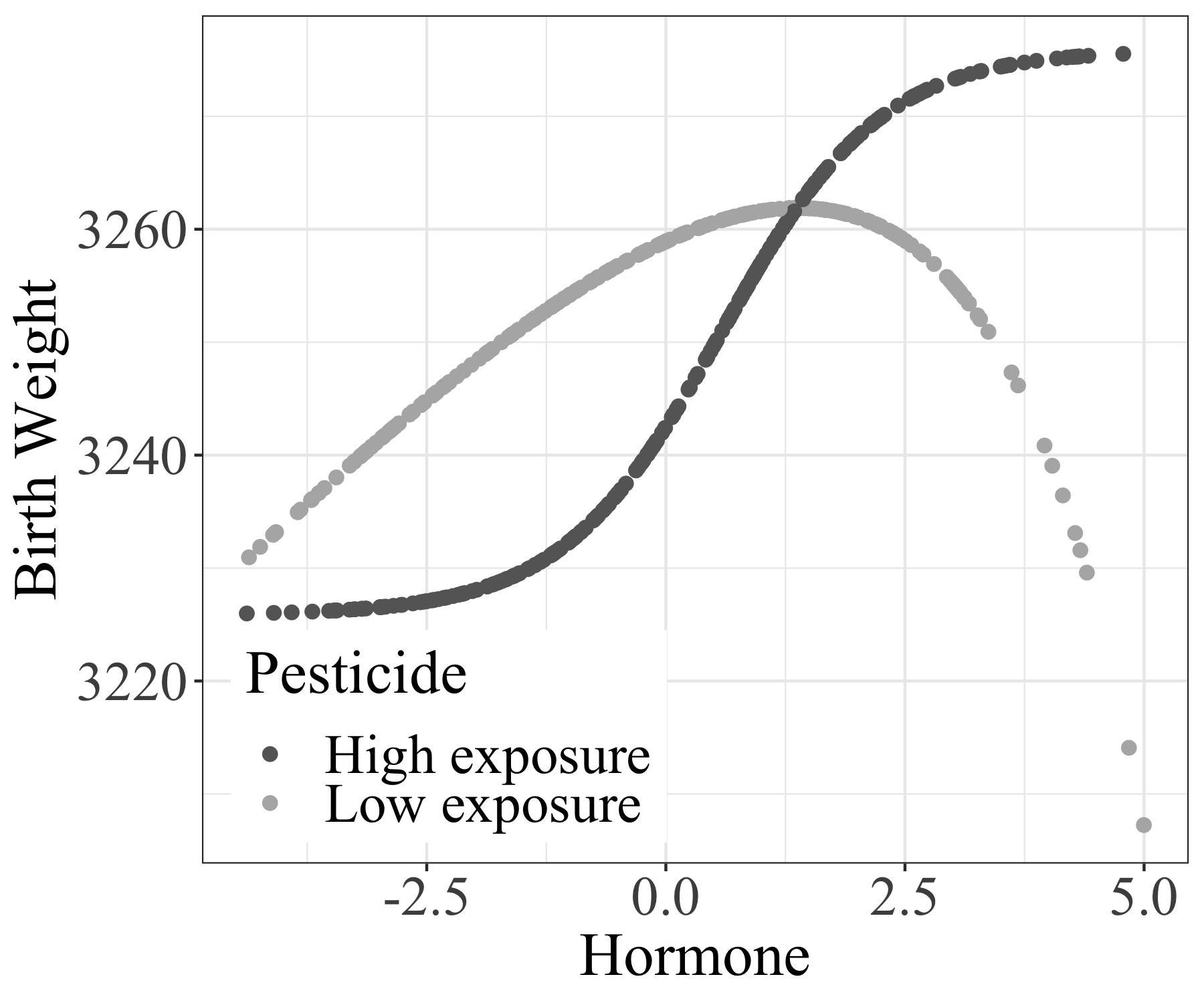}}\hspace{1.5cm}
\subfloat[Coverage Probability under pattern 3]{\label{fig:3:3.b}\includegraphics[width=.4\linewidth]{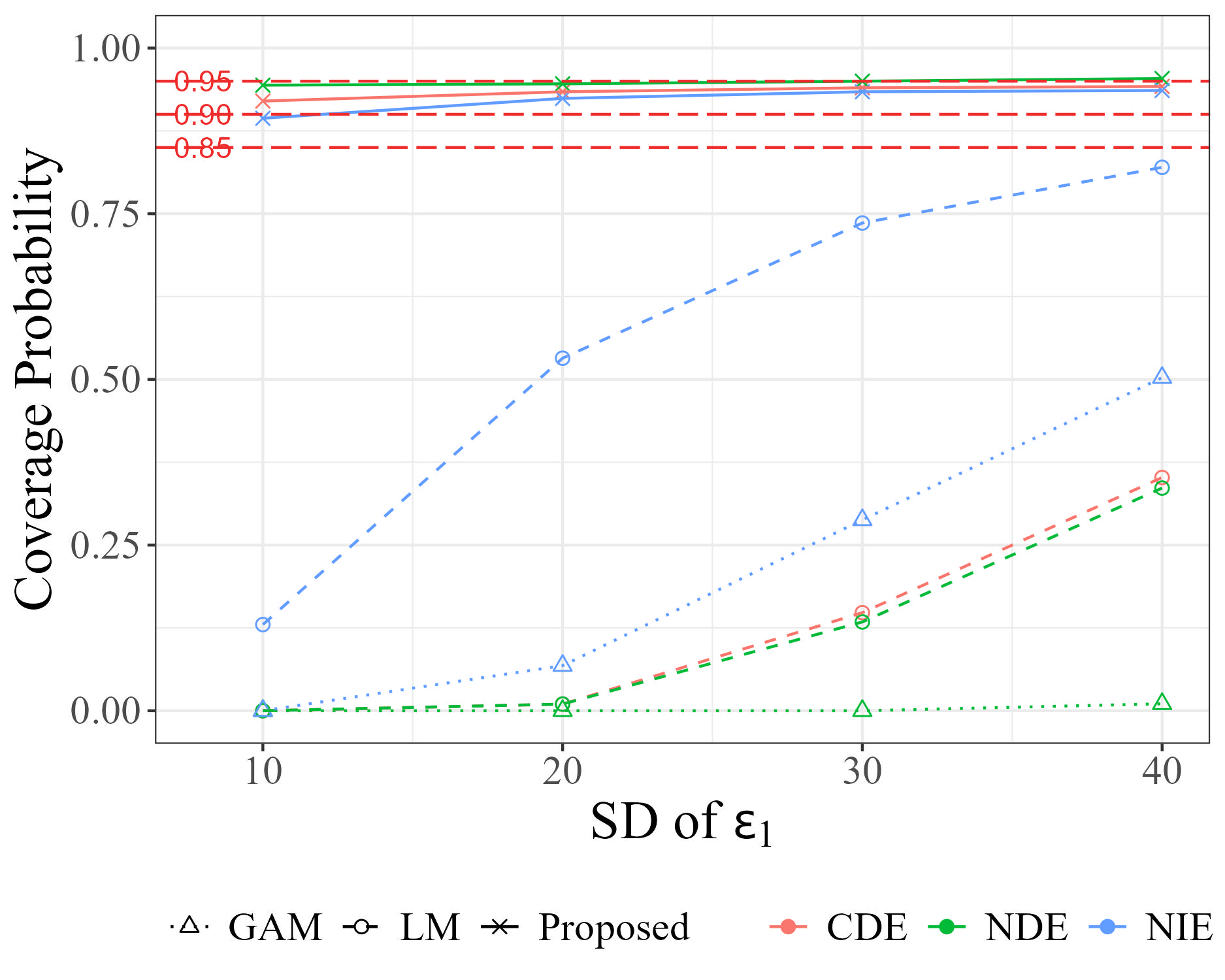}}\par
\caption{Plots of hormone vs. birth weight under each pattern, and simulation results of coverage probability for each pattern }
\label{fig:3:1-3.a-b}
\end{figure}

We consider 3 different combinations of nonlinear functions for exposure-outcome model, which are shown in Figures \ref{fig:3:1.a},  \ref{fig:3:2.a} and \ref{fig:3:3.a}. We fix the variance of $\epsilon_2$ in exposure-mediator model (model (\ref{eq:2.2:2})) at $0.3^2$, and four different patterns of variances are considered for $\epsilon_1$ in exposure-outcome model (model (\ref{eq:2.2:1})), namely, $N(0, 10^2)$, $N(0, 20^2)$, $N(0, 30^2)$ and $N(0, 40^2)$. The number of bases is set to 5. For the CDE, the mediator is set to its mean value. Since in simulation studies all parameters such as $\beta_1$, $\gamma_0$, $\gamma_1$, $\gamma_2$ and $\sigma_2^2$ and the functions $f_1(m)$ and $f_2(m)$ are known, the true effects can be calculated using the formulas in Proposition 3 to investigate the performance of the proposed methodology. The number of simulations considered is 500. The results of coverage probability are shown in Figures \ref{fig:3:1.b},  \ref{fig:3:2.b} and \ref{fig:3:3.b} (``LM'' in the legend of each figure corresponds to the linear regression-based method and ``GAM'' in the legend of each figure corresponds to the GAM (simulation)-based method), and the results of average absolute relative bias and average MSE are shown in Figure S.1 in the supplementary text (also see Tables S.1 - S.3 in the supplementary text for details).

Under Pattern 1 (Figure \ref{fig:3:1.a}), the relationship between hormone and birth weight in the high exposure group is concave (with increasing trend) and the relationship in the low exposure group is increasing (sigmoid). The true CDE is centered at 44.62 (SD = 0.14), the true NDE is 45.82 and the true NIE is 1.10. The coverage probabilities of CDE, NDE and NIE remain at around 95\% for the proposed method under different $\sigma_1$'s, but the coverage probabilities of CDE and NDE are low for the linear regression-based method and the coverage probabilities of NDE are low for the GAM (simulation)-based method, especially in the cases of small $\sigma_1$'s. The average absolute relative biases and the average MSEs for the proposed method are much lower than those for the linear regression-based or GAM (simulation)-based method. Under Pattern 2 (Figure \ref{fig:3:2.a}), the relationship between hormone and birth weight in the high exposure group is concave and the relationship in the low exposure group is concave (with decreasing trend). The true CDE is centered at 19.85 (SD = 0.05), the true NDE is 19.17 and the true NIE is -0.17. The coverage probabilities of CDE, NDE and NIE are approximately 95\% for the proposed method. Although the coverage probabilities of the CDE and NDE for the linear regression-based method and the coverage probabilities of NDE for the GAM (simulation)-based method under Pattern 2 performed better than those under Pattern 1, they are still low in the cases of small $\sigma_1$'s. The average absolute relative biases for all methods converged as $\sigma_1$ increases. The average MSEs of CDE and NDE for the linear regression-based method and the average MSEs of NDE for the GAM (simulation)-based method were lower under large $\sigma_1$'s. The reason could be that the estimations of effects do not deviate much from the truth for the linear regression-based or GAM (simulation)-based method. Under Pattern 3 (Figure \ref{fig:3:3.a}), the relationship between hormone and birth weight in the high exposure group is increasing (sigmoid) and the relationship in the low exposure group is concave. The true CDE is centered at -15.00 (SD = 0.14), the true NDE is -16.24 and the true NIE is 4.08. The coverage probabilities of CDE, NDE and NIE remain at approximately 95\% for the proposed method under different $\sigma_1$'s, but the coverage probabilities of CDE, NDE and NIE are low for the linear regression-based method and the coverage probabilities of NDE and NIE are low for the GAM (simulation)-based method, especially in the cases of small $\sigma_1$'s. The average absolute relative biases and the average MSEs for the proposed method are notably lower than those for the linear regression-based method. The true NIE under Pattern 3 is larger than the true NIEs under other patterns in our simulation study. For this reason, the performance measures of NIE for the linear regression-based or GAM (simulation)-based method are worse. 

In summary, if $f_1(M)$ and $f_2(M)$ deviate from linear shapes, the semi-parametric shape-restricted regression spline outperforms the linear regression or GAM (simulation) in general. Using the semi-parametric shape-restricted regression spline, the coverage probability stays constant at around 95\%; whereas using the linear regression or GAM (simulation), the coverage probability tends to be 0 when the variance of $\epsilon_1$ is small and the effect size is large (see Proposition \ref{prop:3:1}). Both average absolute relative bias and average MSE of the estimated effects from the semi-parametric shape-restricted regression spline increase when the variance of $\epsilon_1$ increase, but they are still smaller compared to the linear regression or GAM (simulation), especially for small variance of $\epsilon_1$ and large effect size. 

For a fair comparison, we have also assumed the true relationships between hormone and birth weight in high and low exposure groups as linear, i.e., $f_1(M)=5.5M + 70$ and $f_2(M)=9.5M + 60$, both increasing (see Figure 5a in the supplementary text). The true CDE is centered at 25.44 (SD = 0.05), the true NDE is 26.07, and the true NIE is 1.65. Because the underlying relationships are linear, the linear regression-based method performed better, as expected. The metrics are quite comparable to those from the semi-parametric shape-restricted regression spline (see Figure S.2 and Table S.4 in the supplementary text).

\begin{proposition} \label{prop:3:1}
If the mediation effect based on the linear regression model is different from the true mediation effect, then as the variance of $\epsilon_1$ $\rightarrow 0$, the coverage probability (or the probability that the confidence interval contains the true mediation effect)  $\rightarrow 0$.
\end{proposition}

The proof of Proposition \ref{prop:3:1} can be found in the supplementary text.
\section{Application}
\label{sec:4}
Pesticides are used widely in agriculture to control pests and improve yields. By design, they are toxic to insects and therefore we and others hypothesize unintended negative consequences on human health \citep{larsen2017}. \cite{larsen2017} reported that for individuals in a high pesticide exposure group (defined by pounds of pesticides applied in residential vicinity) in $1^{st}$-trimester pregnancies, birth weight was 13 grams lower. Being in the high exposure group was associated with lower gestational age, higher risk of preterm birth and higher probability of a birth abnormality. Chemicals may not directly reach and/or affect the fetuses but instead they alter the levels of placental biomarkers which influence the birth outcomes. This is a placentally-mediated effect \citep{adibi2021b}. It is also possible that the environmental chemical is associated with the outcome by way of a nonlinear relationship. Once in the body, some chemicals mimic naturally occurring hormones, prostaglandins, and/or growth factors \citep{leMaire2010}. To evaluate this, we performed mediation analysis using the semi-parametric shape-restricted regression analysis developed in this paper.

\begin{figure}
\centering
\includegraphics[width=0.9\textwidth]{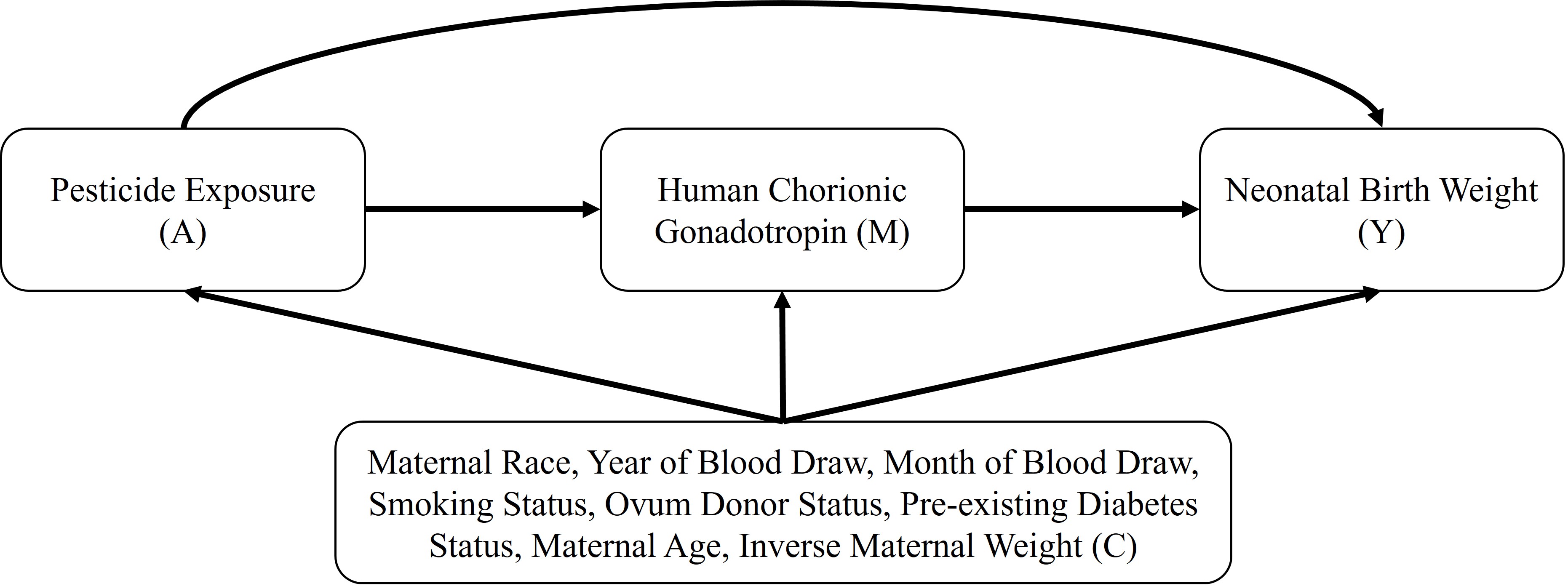}
\caption{The directed acyclic graph (DAG) for the analysis}
\label{fig:4:1}
\end{figure}

Pesticide data were publicly available through the California’s Pesticide Use Reporting (PUR) program, where the pesticide application is reported daily. The data were mapped to the zip code of maternal residence, and to the month and year of blood draw. Each woman was assigned pesticide exposure values in log pounds of pesticides, applied in that zip code during the month in which she had her blood drawn in the first trimester of pregnancy (10-14 weeks gestation). These data were merged with the prenatal screening data in which each woman had precise and accurate measures of placental-fetal hormones. Human chorionic gonadotropin (hCG) was selected as a mediator for this analysis as it is one of five placental-fetal biomarkers that is used widely to screen for fetal aneuploidy and therefore is available in the medical record for research \citep{Malone2005}. hCG is a placental glycoprotein that has been related to almost all placental functions including the transfer of nutrition, fetal growth and development \citep{Filicori2005, Licht2007}. hCG has also been widely associated with environmental exposures including pesticides \citep{Paulesu2018, adibi2021c}. Furthermore, \cite{barjaktarovic2017} demonstrated in Generation R that hCG in the late first trimester was associated with birth weight in a sex-specific and nonlinear pattern. The pesticide application data was dichotomized at the median to represent low and high exposure categories ($A$). We focused on two commonly used pesticides, namely, permethrin and glyphosate isopropylamine salt. The mediator ($M$) was her serum level human chorionic gonadotropin (hCG), normalized for gestational day of blood draw which is called the gestational age multiple of the median (GA-MoM) \citep{adibi2015b}. The outcome variable ($Y$) was neonatal birth weight which was abstracted from the medical record for each baby. Analyses were stratified by baby sex due to previous reports on a sex difference in the hCG-birth weight association \citep{barjaktarovic2017}. All models were adjusted for a small set of confounders of exposure-outcome, exposure-mediator, and mediator-outcome that included maternal race, year of blood draw, month of blood draw, smoking status, ovum donor status, pre-existing diabetes status, maternal age and inverse maternal weight \citep{adibi2015b}. Information on confounders was limited to a one-page questionnaire which is completed by subjects at the time of blood draw. The DAG is shown in Figure \ref{fig:4:1}.

For the organophosphate pesticide permethrin ($N_{male} = 21,433, N_{female} = 20,895$), the association between first trimester hCG and neonatal male birth weight was increasing in the above-median exposure group and concave with an increasing trend in the below-median exposure group. The relationship between first trimester hCG and neonatal female birth weight was concave with an increasing trend in the above-median exposure group and increasing in the below-median exposure group. For glyphosate isopropylamine salt ($N_{male} = 56,299, N_{female} = 55,052$), the relationship between first trimester hCG and infant male birth weight was concave with an increasing trend in the above-median exposure group and increasing in the below-median exposure group. The relationship between first trimester hCG and infant female birth weight was increasing in the above-median exposure group and concave with increasing trend in the below-median exposure group. Mediation analysis was performed using the semi-parametric shape-restricted regression spline. The number of bases was set to 5 as in the simulation study and the confounding variables were controlled at their mean values. For the CDE, the mediator was set to its mean value. The results are summarized in Table \ref{tab:4.2:1}.

\begin{table}
\begin{center}
\caption{Mediation Analysis Results on State-Wide Prenatal Screening Program Data}
\label{tab:4.2:1}
\begin{tabular}{ c c c c } 
\hline
\multicolumn{4}{c}{Female Infant}\\
\hline
 & CDE with 95\% C.I. & NDE with 95\% C.I. & NIE with 95\% C.I. \\
\hline
permethrin & \makecell{13.270 \\ (-4.0927, 30.633)} & \makecell{6.7042 \\ (-5.2455, 18.654)} & \makecell{-1.1889 \\ (-2.0636, -0.3142)} \\
\hline
\makecell{glyphosate \\ isopropylamine salt} & \makecell{4.9524 \\ (-6.9646, 16.869)} & \makecell{8.8868 \\ (-11.683, 29.457)} & \makecell{-1.6303 \\ (-2.2427, -1.0178)} \\
\hline \hline
\multicolumn{4}{c}{Male Infant}\\
\hline
 & CDE with 95\% C.I. & NDE with 95\% C.I. & NIE with 95\% C.I. \\
\hline
permethrin & \makecell{12.454 \\ (-3.5468, 28.455)} & \makecell{15.665 \\ (3.5325, 27.797)} & \makecell{-2.4189 \\ (-3.5479, -1.2899)} \\
\hline
\makecell{glyphosate \\ isopropylamine salt} & \makecell{8.0234 \\ (-3.0360, 19.083)} & \makecell{5.8625 \\ (-1.6837, 13.409)} & \makecell{-2.4338 \\ (-3.1993, -1.6683)} \\
\hline
\end{tabular}
\end{center}
\end{table}

According to the results summarized in Table \ref{tab:4.2:1}, the NIE is statistically significant at 5\% level of significance in each case as the 95\% confidence intervals do not include the null value, indicating that first trimester hCG was in an association pathway between permethrin and glyphosate isopropylamine salt exposure and birth weight within male and female infants in this cohort. In our analysis, we discovered positive associations between permethrin and birth weight and between glyphosate isopropylamine salt and birth weight. After accounting for mediation, these associations became negative. It is important to note that pesticide exposure is based on application of pesticides and geographic proximity. Results can be interpreted at the aggregate level of women living in geographic proximity to specific levels of pesticide application in early pregnancy. At the individual level, misclassification of internal pesticide dose is likely.
\section{Discussion}
\label{sec:5}
The methodology presented here solves a problem in the setting of causal mediation analysis where there is a known or suspected nonlinear relationship between the mediator and the outcome. This method is designed to relax the linearity assumption and reduce bias introduced by model misspecification. The exposure-outcome model is specified using the quadratic I-spline basis functions and/or cubic C-spline basis functions. This allows the investigator to apply pre-existing knowledge on the underlying nonlinear relationship to their analysis. Once the shapes of the associations are established, the mediation effects can be estimated and inferred.

We demonstrate the proposed method in an applied example in environmental health and fetal origins epidemiology. In the case of placental-fetal biomarkers and fetal growth, nonlinearity is a reasonable assumption. We found that the shapes of the placental biomarker-birth weight relationships differed by the low and high pesticide application groups. One interpretation is that the shape of association in the low exposure group more closely represents the ``normal'' developmental relationship, and the changing of that shape in the high exposure group could indicate a type of toxicity. This information was accounted for in the estimation of the controlled direct effect (the exposure effect when hypothetically blocking the effect of the mediator), the natural direct effect (the exposure effect when assuming the mediator was unaffected by exposure), and the natural indirect effect (the effect of exposure assuming mediation). In this case, it allowed for comparisons of effect magnitude, direction, and precision of the effect across different pesticides and hormones, stratified by sex of the infant. 

As shown in the Application section (Section \ref{sec:4}), the estimation of the natural indirect effect changed the interpretation of the exposure effect. Based on the natural direct effect alone, the conclusion was that the pesticide application was either not or positively associated with the birth weight. However, based on the natural indirect effect and the consideration of the shapes, the conclusion was that the increased pesticide application would lead to the lower birth weight. This question as to whether the true exposure effect is positive or negative in direction can be further explored in studies where exposures are measured in individual pregnancies by pesticide biomarkers. A possible explanation is that the placental mechanism of toxicity differs from the direct effect, and both directions of association (positive natural direct effect and negative natural indirect effect) may be accurate. It is also important to note that, if the relationships are modeled using standard linear models, then from equations $E[Y_{aM_{a^*}}-Y_{a^*M_{a^*}}|c] = (\beta_1 + \beta_3 (\gamma_0 + \gamma_1 a^* + \gamma_2 c))(a - a^*)$ and $E[Y_{aM_a}-Y_{aM_{a^*}}|c] = (\beta_2 \gamma_1 + \beta_3 \gamma_1 a)(a - a^*)$ \citep{vanderweele2015}, we see that under some conditions on the model parameters, $E(NDE) > 0 > E(NIE)$. This can happen, for example, when in the exposure-outcome model, the exposure, the mediator as well as the interaction between the exposure and the mediator are positively correlated with the outcome. Conversely, in the exposure-mediator model, the confounders and the mediator are positively correlated while the exposure and the mediator are negatively correlated. There are potentially other configurations leading to $E(NDE) > 0 > E(NIE)$. Thus, it should not be surprising if in practice, whether we use the linear or the proposed shape-restricted regression, the natural direct effect and the natural indirect effect have opposite signs as seen in our example. 

Using these estimation methods in the causal inference framework requires consideration of specific sources of bias and rigorous validation of the necessary assumptions including no unmeasured confounding, consistency, positivity, and no interference. Correct model specification is a required assumption in this framework, which is addressed in this method. Detailed discussion of these assumptions and approaches to validate them are outside the scope of this paper and are described in detail elsewhere \citep{hernan2020, nguyen2021}.

The proposed method performed as desired in terms of coverage probability, average absolute relative bias and average mean squared error based on the simulation results. This was true especially when the underlying shape was nonlinear with a specific shape being identified. Standard linear or polynomial regressions are rigid in shapes, so the corresponding 95\% confidence intervals barely covered the true effect. The proposed method also avoids the common problems in GAMs, i.e., the curse of dimensionality, computational intensity, etc. Due to the flexibility offered by GAMs, the associations may be unstable and interpretation can become challenging. Because the proposed method is developed using the asymptotic properties of the regression, it is not computationally intensive. The proposed method is a compromise between the standard linear or polynomial regressions and GAMs because it does not require a specific functional relationship nor does it allow arbitrary relationships. Although the fits are robust to the choice of smoothing parameters when assumptions about both shape and smoothness are warranted \citep{meyer2008}, in order to make more precise predictions, the number of knots and the knots placement may still need to be considered. See for example, \cite{jelsema2019}. The proposed method is not suitable if shapes other than monotonic, convex and concave are present. In such cases, researchers can apply the simulation-based method developed by \cite{imai2010}. The proposed method only accommodates a categorical exposure variable. This is a limitation to be addressed in the future extension of this approach.
\section*{Acknowledgements}

The authors QY and JJA gratefully acknowledge the funding from the National Institute of Environmental Health Sciences (NIEHS 5R01ES029336). The research of SDP was funded by the Intramural Research Program of the {\it Eunice Kennedy Shriver} National Institute of Child Health and Human Development (NICHD), National Institutes of Health, Bethesda, MD. We acknowledge Drs. Sona Saha and Robert Currier of the Genetic Disease Screening Program (GDSP) of the California Department of Public Health (CDPH) for their support of this collaboration. We thank Dr. Jane Clougherty (Drexel University) and Ms. Ellen Kinnee (University of Pittsburgh) for their work to assign pesticide application exposures (PUR data) to individual pregnancies (GDSP data) based on different geo-spatial units. We thank Dr. Jiebiao Wang (University of Pittsburgh) for his comments and help on this paper.

\bibliography{etdbib}

\end{document}


\renewcommand\thefigure{S.\arabic{figure}} 
\renewcommand\thetable{S.\arabic{table}}
\renewcommand\theequation{S.\arabic{equation}}
\section*{Supplementary Text}
\subsection*{S.1. M-splines, I-splines and C-splines}

\noindent
Let $t = \{t_1, t_2, ..., t_{n+k}\}$ denote the knot sequence, $n$ denote the number of free parameters that specify the spline function having the specified continuity characteristics, and $k$ be the order of the basis functions. Then the recursive form of M-splines is as follows (Ramsay, 1988): 

\noindent
For order $k = 1$, $M_i(x|1, t) = \frac{1}{t_{i+1}-t_i}$ if $t_i \leq x < t_{i+1}$, otherwise $M_i(x|1, t) = 0$, and 

\noindent
for order $k > 1$, $M_i(x|k,t) = \frac{k[(x-t_i)M_i(x|k-1,t) + (t_{i+k}-x)M_{i+1}(x|k-1,t)]}{(k-1)(t_{i+k}-t_i)}$ if $t_i \leq x < t_{i+k}$, otherwise $M_i(x|k,t) = 0$.

\bigskip
\noindent
The quadratic I-splines $I_i(x|2,t)$ are obtained by integrating the M-splines of degree 1 and can be expressed as: 

\noindent
$I_i(x|2,t) = 0$ if $x < t_i$, 

\noindent
$I_i(x|2,t) = \frac{(x-t_i)^2}{(t_{i+2}-t_{i})(t_{i+1}-t_i)}$ if $t_i \leq x < t_{i+1}$, 

\noindent
$I_i(x|2,t) = 1 - \frac{(t_{i+2}-x)^2}{(t_{i+2}-t_{i})(t_{i+2}-t_{i+1})}$ if $t_{i+1} \leq x < t_{i+2}$, and

\noindent
$I_i(x|2,t) = 1$ if $x \geq t_{i+2}$.

\bigskip
\noindent
The cubic C-splines $C_i(x|2,t)$ are obtained by integrating the quadratic I-splines and can be expressed as: 

\noindent
$C_i(x|2,t) = 0$ if $x < t_i$, 

\noindent
$C_i(x|2,t) = \frac{(x-t_i)^3}{3(t_{i+2}-t_{i})(t_{i+1}-t_i)}$ if $t_i \leq x < t_{i+1}$, 

\noindent
$C_i(x|2,t) = x - \frac{t_i+t_{i+1}+t_{i+2}}{3} + \frac{(t_{i+2}-x)^3}{3(t_{i+2}-t_{i})(t_{i+2}-t_{i+1})}$ if $t_{i+1} \leq x < t_{i+2}$, and

\noindent
$C_i(x|2,t) = x - \frac{t_i+t_{i+1}+t_{i+2}}{3}$ if $x \geq t_{i+2}$.

\bigskip
\subsection*{S.2. Proof of Proposition 1}
\begin{proof}
Under the conditions described in Proposition 1, and with the models (1) and (2) being correctly specified, we obtain after some simplifications:

\noindent
$E[CDE|c] = E[Y_{am} - Y_{a^*m}|c] = E[Y|a, m, c] - E[Y|a^*, m, c]  = (\beta_1 + f_1(m) - f_2(m))(a - a^*)$,

\noindent
$E[NDE|c] = E[Y_{aM_{a^*}}-Y_{a^*M_{a^*}}|c] = \int_m \{E[Y|a, m, c] - E[Y|a^*, m, c]\}f(m|a^*, c) dm$ 

\noindent
$= (\beta_1 + E[f_1(M)|a^*, c] - E[f_2(M)|a^*, c])(a - a^*)$, and

\noindent
$E[NIE|c] = E[Y_{aM_a}-Y_{aM_{a^*}}|c] = \int_m E[Y|a, m, c]\{f(m|a, c) - f(m|a^*, c)\} dm$ 

\noindent
$= a(E[f_1(M)|a, c] - E[f_1(M)|a^*, c]) + (1 - a)(E[f_2(M)|a, c] - E[f_2(M)|a^*, c])$.

\bigskip
\noindent
Let the knot sequence be $L = t_1 = t_2 < t_3 < ... < t_k < t_{k+1} = t_{k+2} = U$. If $f_1(M)$ is fitted using I-splines, then $f_1(M) = \beta_{21} I_1(M|2, t) + ... + \beta_{2k} I_k(M|2, t)$. This is a piece-wise function, where, for $t_k \leq M < t_{k+1}$, $f_1(M) = \beta_{21} + ... + \beta_{2, k-2} + \beta_{2, k-1}(1-\frac{(t_{k+1}-M)^2}{(t_{k+1}-t_k)(t_{k+1}-t_{k-1})}) + \beta_{2, k}(\frac{(M-t_k)^2}{(t_{k+1}-t_k)(t_{k+2}-t_k)})$. Then conditional on $a$ and $c$ we obtain: 

\bigskip
\noindent
$E[f_1(M)|a, c] = \int_m (\beta_{21} I_1(m|2, t) + ... + \beta_{2k} I_k(m|2, t))f(m|a, c) dm = \sum_{i=2}^k \{\int_{t_i}^{t_{i+1}} [\beta_{21} + ... + \beta_{2, i-2} + \beta_{2, i-1}(1-\frac{(t_{i+1}-m)^2}{(t_{i+1}-t_i)(t_{i+1}-t_{i-1})}) + \beta_{2, i}(\frac{(m-t_i)^2}{(t_{i+1}-t_i)(t_{i+2}-t_i)})]f(m|a, c)dm\}$, where $f(m|a, c)$ denotes normal density with mean $\gamma_0+\gamma_1 a+\gamma_2 c$ and variance $\sigma_2^2$.  Similar expressions can be derived for $f_2(M)$ and $E[f_2(M)|a, c]$. 

\bigskip
\noindent
If $f_1(M)$ is fitted using C-splines, then $f_1(M) = \beta_{20} M + \beta_{21} C_1(M|2, t) + ... + \beta_{2k} C_k(M|2, t)$, which is also a piece-wise function. For either $M < t_2$ or $M \geq t_{k+1}$,   $f_1(M) = \beta_{20} M$. For $t_k \leq M < t_{k+1}$,  $f_1(M) = \beta_{20} M + \beta_{21}(M - \frac{t_1+t_2+t_3}{3}) + ... + \beta_{2, k-2}(M - \frac{t_{k-2}+t_{k-1}+t_{k}}{3}) + \beta_{2, k-1}(m - \frac{t_{k-1}+t_k+t_{k+1}}{3} + \frac{(t_{k+1}-m)^3}{3(t_{k+1}-t_k)(t_{k+1}-t_{k-1})}) + \beta_{2, k}(\frac{(m-t_k)^3}{3(t_{k+1}-t_k)(t_{k+2}-t_k)})$. Then conditional on $a$ and $c$ we obtain:

\bigskip
\noindent
$E[f_1(M)|a, c] = \int_m (\beta_{20} m + \beta_{21} C_1(m|2, t) + ... + \beta_{2k} C_k(m|2, t))f(m|a, c)dm =$
 
\noindent
$\beta_{20}(\gamma_0 + \gamma_1 a + \gamma_2 c) + \sum_{i=2}^k \{\int_{t_i}^{t_{i+1}} [\beta_{21}(m - \frac{t_1+t_2+t_3}{3}) + ... + \beta_{2, i-2}(m - \frac{t_{i-2}+t_{i-1}+t_i}{3})$
 
\noindent
$+ \beta_{2, i-1}(m - \frac{t_{i-1}+t_i+t_{i+1}}{3} + \frac{(t_{i+1}-m)^3}{3(t_{i+1}-t_i)(t_{i+1}-t_{i-1})}) + \beta_{2, i}(\frac{(m-t_i)^3}{3(t_{i+1}-t_i)(t_{i+2}-t_i)})]f(m|a, c)dm\}$.
 
\bigskip
\noindent
Similar expressions can be derived for $f_2(M)$ and $E[f_2(M)|a, c]$.
\end{proof}

\subsection*{S.3. Proposition 3}
\begin{customprop}{3}
Let $\theta_{CDE} = (\beta_1, \beta_2, \beta_3)$, $\theta_{NDE} = (\beta_1, \beta_2, \beta_3, \gamma_0, \gamma_1, \gamma_2, \sigma_2^2)$ and $\theta_{NIE} = (\beta_2, \gamma_0, \gamma_1, \gamma_2, \sigma_2^2)$, where $\beta_2 = (\beta_{21}, ..., \beta_{2k})$ and $\beta_3 = (\beta_{31}, ..., \beta_{3k})$ in case of quadratic I-splines, and $\beta_2 = (\beta_{20}, \beta_{21}, ..., \beta_{2k})$ and $\beta_3 = (\beta_{30}, \beta_{31}, ..., \beta_{3k})$ in case of cubic C-splines. Denote the expected controlled direct effect as $g_{CDE}(\theta_{CDE})$, the expected natural direct effect as $g_{NDE}(\theta_{NDE})$ and the expected natural indirect effect as $g_{NIE}(\theta_{NIE})$. Then the asymptotic variances of expected CDE, NDE and NIE are $\nabla_{\theta_{CDE}} g_{CDE}(\theta_{CDE})^T \Sigma_{\theta_{CDE}} \nabla_{\theta_{CDE}} g_{CDE}(\theta_{CDE})$, \\ $\nabla_{\theta_{NDE}} g_{NDE}(\theta_{NDE})^T \Sigma_{\theta_{NDE}} \nabla_{\theta_{NDE}} g_{NDE}(\theta_{NDE})$ and $\nabla_{\theta_{NIE}} g_{NIE}(\theta_{NIE})^T \Sigma_{\theta_{NIE}} \nabla_{\theta_{NIE}} g_{NIE}(\theta_{NIE})$ respectively, where $\Sigma_{\theta_{CDE}}$, $\Sigma_{\theta_{NDE}}$ and $\Sigma_{\theta_{NIE}}$ are the covariance matrices corresponding to the estimated $\theta_{CDE}$, $\theta_{NDE}$ and $\theta_{NIE}$. 
\end{customprop}

\noindent
If $f_1(M)$ is fitted using I-splines, then:

\bigskip
\noindent
$\frac{\partial E[f_1(M)|a, c]}{\partial \beta_{2i}} = \int_{t_i}^{t_{i+1}} \frac{(m-t_i)^2}{(t_{i+1}-t_i)(t_{i+2}-t_i)}f(m|a, c)dm + \int_{t_{i+1}}^{t_{i+2}} (1-\frac{(t_{i+2}-m)^2}{(t_{i+2}-t_{i+1})(t_{i+2}-t_i)})f(m|a, c)dm + \int_{t_{i+2}}^{t_{i+3}} f(m|a, c)dm + ... + \int_{t_{k}}^{t_{k+1}} f(m|a, c)dm$, for $i = 1, 2, \dots, k$,

\bigskip
\noindent
$\frac{\partial E[f_1(M)|a, c]}{\partial \gamma_0} = \sum_{i=2}^k \{\int_{t_i}^{t_{i+1}} [\beta_{21} + ... + \beta_{2, i-2} + \beta_{2, i-1}(1-\frac{(t_{i+1}-m)^2}{(t_{i+1}-t_i)(t_{i+1}-t_{i-1})}) + \beta_{2, i}(\frac{(m-t_i)^2}{(t_{i+1}-t_i)(t_{i+2}-t_i)})]$ $f(m|a, c)\frac{2(m - (\gamma_0+\gamma_1 a+\gamma_2 c)}{2\sigma_2^2}dm\}$, 

\bigskip
\noindent
$\frac{\partial E[f_1(M)|a, c]}{\partial \sigma_2^2} = \sum_{i=2}^k \{\int_{t_i}^{t_{i+1}} [\beta_{21} + ... + \beta_{2, i-2} + \beta_{2, i-1}(1-\frac{(t_{i+1}-m)^2}{(t_{i+1}-t_i)(t_{i+1}-t_{i-1})}) +$

\noindent
$\beta_{2, i}(\frac{(m-t_i)^2}{(t_{i+1}-t_i)(t_{i+2}-t_i)})]f(m|a, c)(-\frac{1}{2\sigma_2^2}+\frac{(m-(\gamma_0+\gamma_1 a+\gamma_2 c))^2}{2(\sigma_2^2)^2})dm\}$. 

\bigskip
\noindent
Similar expressions can be derived for $\frac{\partial E[f_1(M)|a, c]}{\partial \gamma_1}$ and $\frac{\partial E[f_1(M)|a, c]}{\partial \gamma_2}$. 

\bigskip
\noindent
If $f_1(M)$ is fitted using C-splines, then:

\bigskip
\noindent
$\frac{\partial E[f_1(M)|a, c]}{\partial \beta_{20}} = \gamma_0 + \gamma_1 a + \gamma_2 c$, 

\bigskip
\noindent
$\frac{\partial E[f_1(M)|a, c]}{\partial \beta_{2i}} = \int_{t_i}^{t_{i+1}} \frac{(m-t_i)^3}{3(t_{i+1}-t_i)(t_{i+2}-t_i)}f(m|a, c)dm + \int_{t_{i+1}}^{t_{i+2}} (m - \frac{t_i+t_{i+1}+t_{i+2}}{3} + \frac{(t_{i+2}-m)^3}{3(t_{i+2}-t_{i+1})(t_{i+2}-t_i)})f(m|a, c)dm + \int_{t_{i+2}}^{t_{i+3}} (m - \frac{t_i+t_{i+1}+t_{i+2}}{3})f(m|a, c)dm + ... + \int_{t_{k}}^{t_{k+1}} (m - \frac{t_i+t_{i+1}+t_{i+2}}{3})f(m|a, c)dm$, for $i = 1, 2, \dots, k$ 

\bigskip
\noindent
$\frac{\partial E[f_1(M)|a, c]}{\partial \gamma_0} = \beta_{20}  + \sum_{i=2}^k \{\int_{t_i}^{t_{i+1}} [\beta_{21}(m - \frac{t_1+t_2+t_3}{3}) + ... + \beta_{2, i-2}(m - \frac{t_{i-2}+t_{i-1}+t_i}{3}) + \beta_{2, i-1}(m - \frac{t_{i-1}+t_i+t_{i+1}}{3} + \frac{(t_{i+1}-m)^3}{3(t_{i+1}-t_i)(t_{i+1}-t_{i-1})}) + \beta_{2, i}(\frac{(m-t_i)^3}{3(t_{i+1}-t_i)(t_{i+2}-t_i)})]f(m|a, c)\frac{2(m - (\gamma_0+\gamma_1 a+\gamma_2 c)}{2\sigma_2^2}dm\}$, 

\bigskip
\noindent
$\frac{\partial E[f_1(M)|a, c]}{\partial \sigma_2^2} = \sum_{i=2}^k \{\int_{t_i}^{t_{i+1}} [\beta_{21}(m - \frac{t_1+t_2+t_3}{3}) + ... + \beta_{2, i-2}(m - \frac{t_{i-2}+t_{i-1}+t_i}{3}) + \beta_{2, i-1}(m - \frac{t_{i-1}+t_i+t_{i+1}}{3} + \frac{(t_{i+1}-m)^3}{3(t_{i+1}-t_i)(t_{i+1}-t_{i-1})}) + \beta_{2, i}(\frac{(m-t_i)^3}{3(t_{i+1}-t_i)(t_{i+2}-t_i)})]f(m|a, c)(-\frac{1}{2\sigma_2^2}+\frac{(m-(\gamma_0+\gamma_1 a+\gamma_2 c))^2}{2(\sigma_2^2)^2})dm\}$.

\bigskip
\noindent
Similarly expressions can be derived for $\frac{\partial E[f_1(M)|a, c]}{\partial \gamma_1}$ and $\frac{\partial E[f_1(M)|a, c]}{\partial \gamma_2}$.

\subsection*{S.4. Proof of Proposition 2}
\begin{proof}
Using linear regression, the exposure-outcome model is expressed as 
\begin{equation} \label{eq:a.4:1}
    Y = \beta_0 + \beta_1 A + \beta_2 M + \beta_3 A M + \beta_4 C + \epsilon_1
\end{equation}
where $\epsilon_1 \sim N(0, \sigma_1^2)$, while the exposure-mediator model keeps the same as model (2). Then the expected CDE, NDE and NIE, conditioning on $C = c$, are given by 
\begin{equation} \label{eq:a.4:2}
    E[Y_{am} - Y_{a^*m}|c] = (\beta_1 + \beta_3 m)(a - a^*),
\end{equation}
\begin{equation} \label{eq:a.4:3}
    E[Y_{aM_{a^*}}-Y_{a^*M_{a^*}}|c] = (\beta_1 + \beta_3 (\gamma_0 + \gamma_1 a^* + \gamma_2 c))(a - a^*),
\end{equation}
and
\begin{equation} \label{eq:a.4:4}
    E[Y_{aM_a}-Y_{aM_{a^*}}|c] = (\beta_2 \gamma_1 + \beta_3 \gamma_1 a)(a - a^*),
\end{equation}
respectively. Let $X = [1, A, M, AM, C]$, then $\hat{\beta} \sim N(\beta, \sigma_1^2 (X^T X)^{-1})$. The expected CDE, NDE and NIE can all be expressed as a linear combination of $\beta$. Assuming  $\hat{\theta}_{LR} = a\hat{\beta} \sim N(a\beta, \sigma_1^2 a(X^T X)^{-1}a^T)$, then we obtain $P(|\hat{\theta}_{LR} - \theta_{true}| \leq z_{\alpha/2} \sqrt{Var(\hat{\theta}_{LR})}) =  \phi(z_{\alpha/2} + \frac{\theta_{true} - \theta_{LR}}{\sigma_1\sqrt{a(X^T X)^{-1}a^T}}) - \phi(-z_{\alpha/2} + \frac{\theta_{true} - \theta_{LR}}{\sigma_1\sqrt{a(X^T X)^{-1}a^T}})$, where $\phi(\cdot)$ denotes normal density function. Therefore, if $\theta_{true} \neq \theta_{LR} $ then as $\sigma_1 \rightarrow 0$,  the coverage probability $\rightarrow 0$.
\end{proof}

\subsection*{S.5. Simulation results of average absolute relative bias and average MSE for each pattern}
\begin{figure}[H]
\centering
\subfloat[Average $|$Relative Bias$|$ under pattern 1]{\label{fig:3:1.c}\includegraphics[width=.45\linewidth]{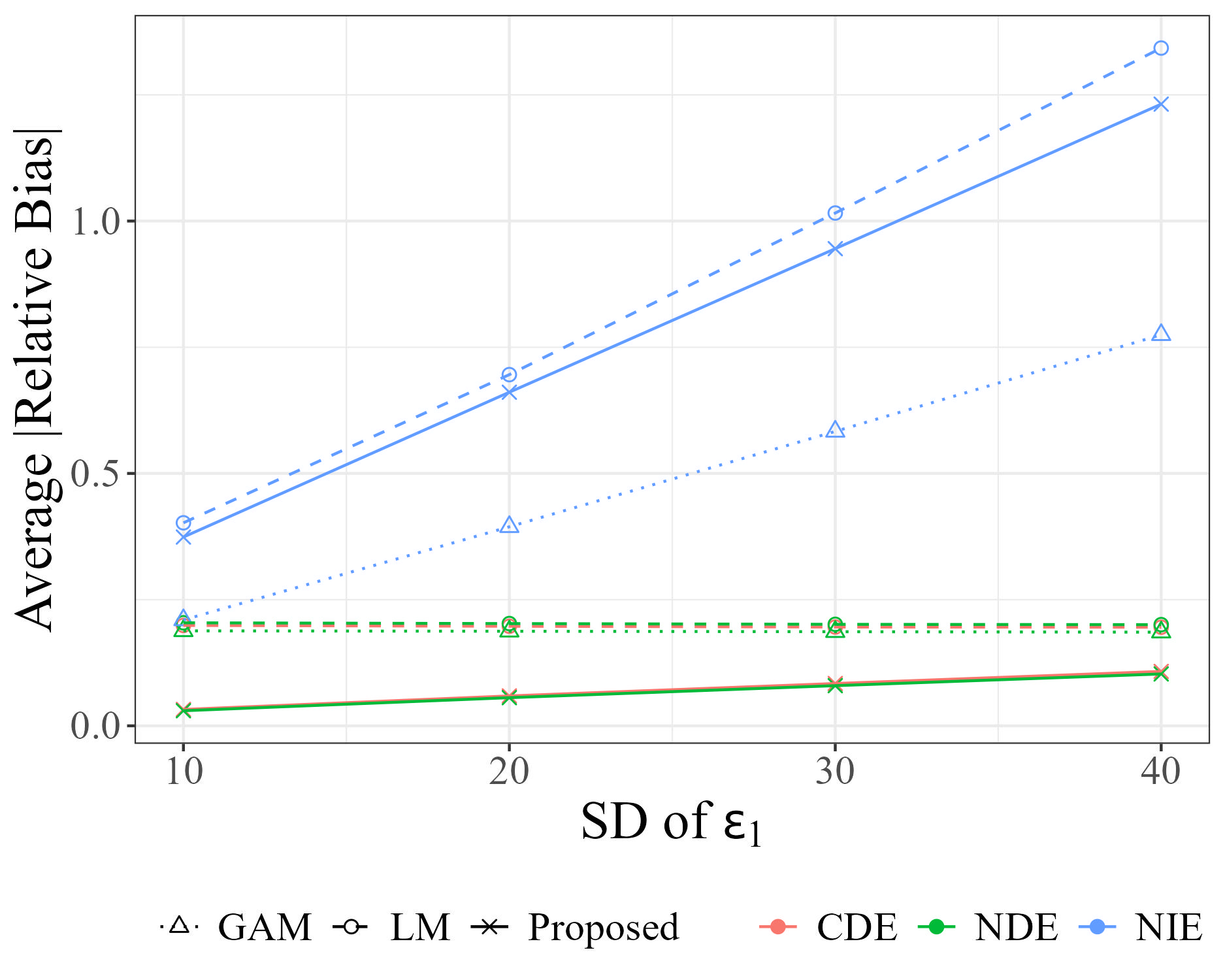}}\hfill
\subfloat[Average MSE under pattern 1]{\label{fig:3:1.d}\includegraphics[width=.45\linewidth]{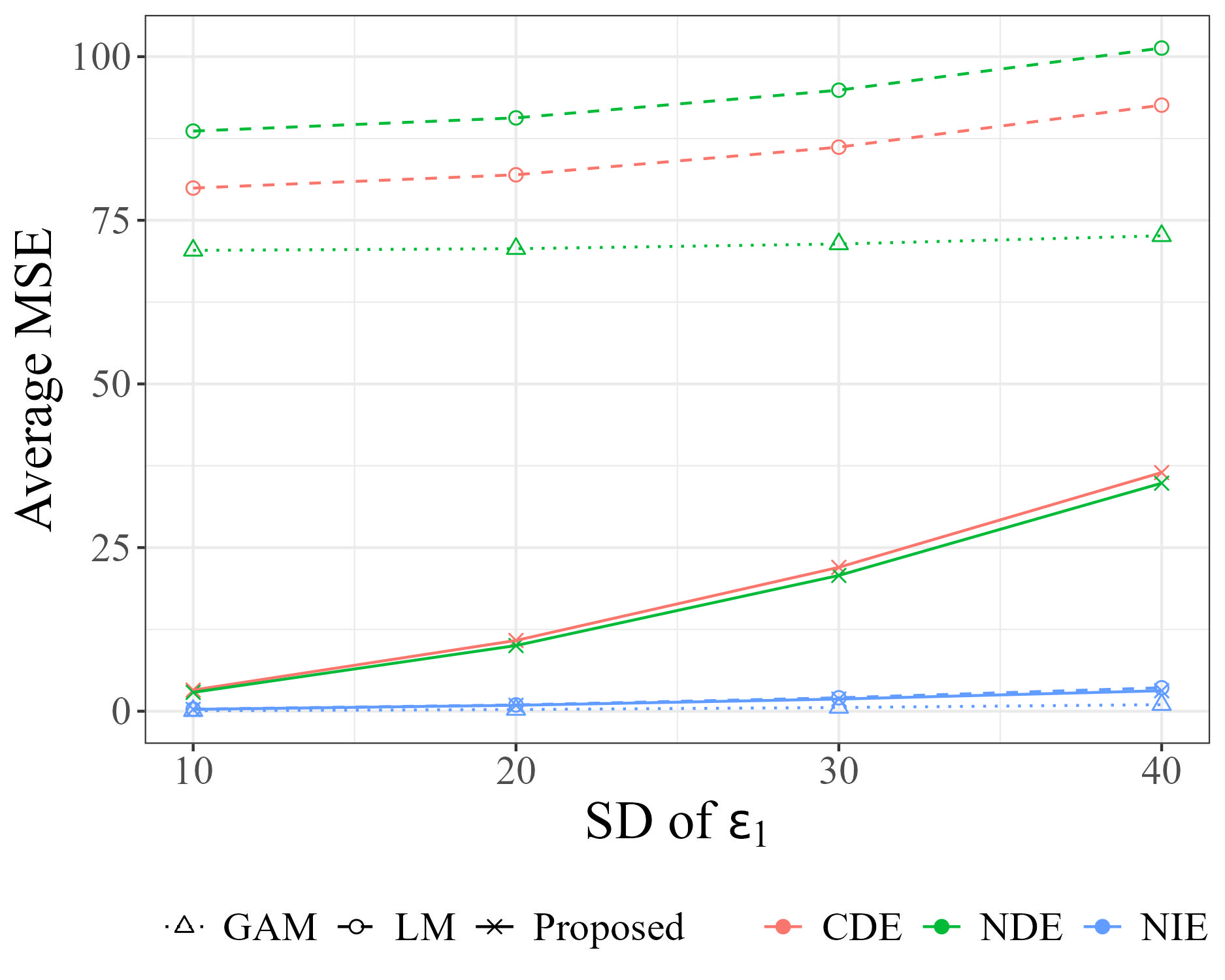}}\par
\subfloat[Average $|$Relative Bias$|$ under pattern 2]{\label{fig:3:2.c}\includegraphics[width=.45\linewidth]{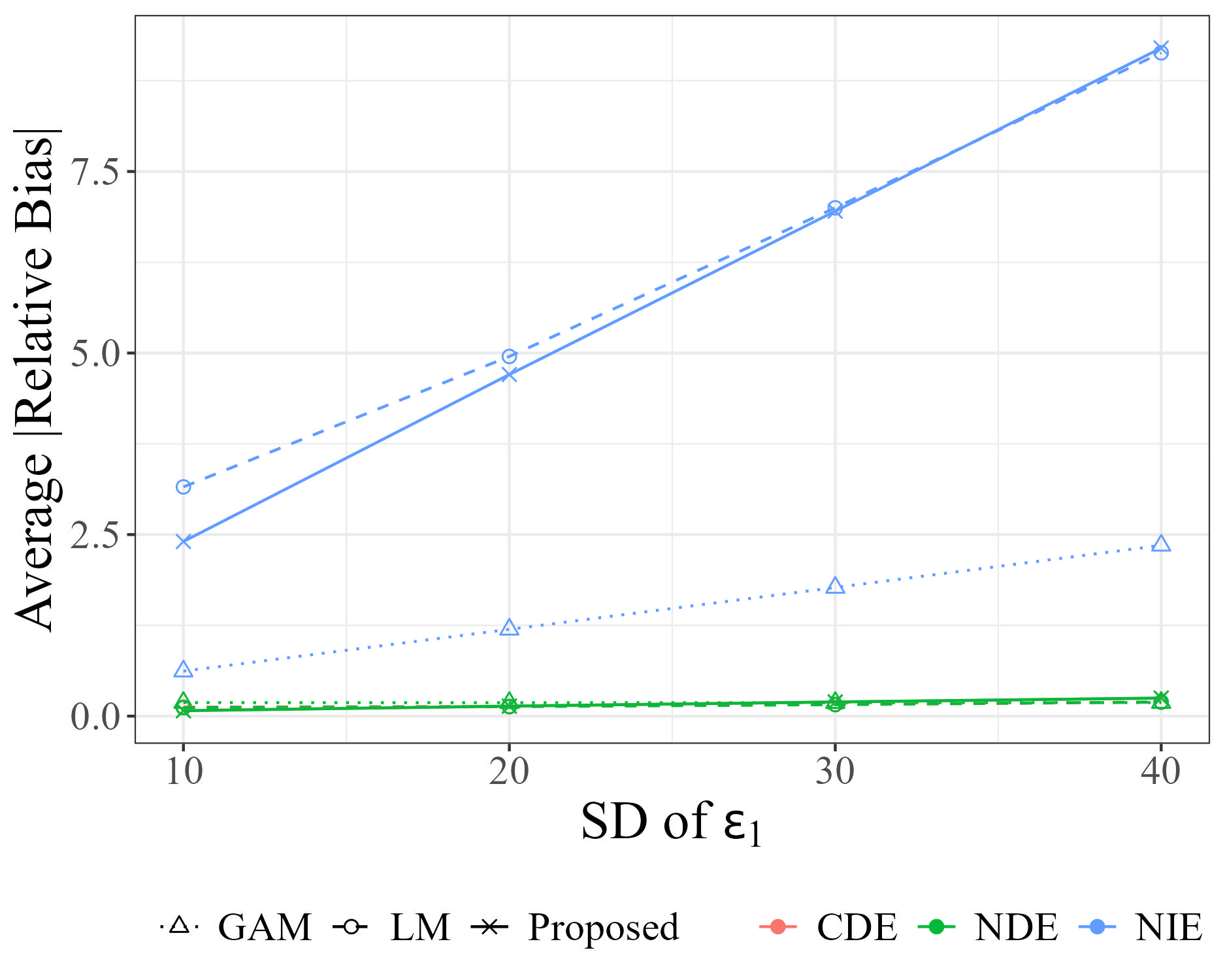}}\hfill
\subfloat[Average MSE under pattern 2]{\label{fig:3:2.d}\includegraphics[width=.45\linewidth]{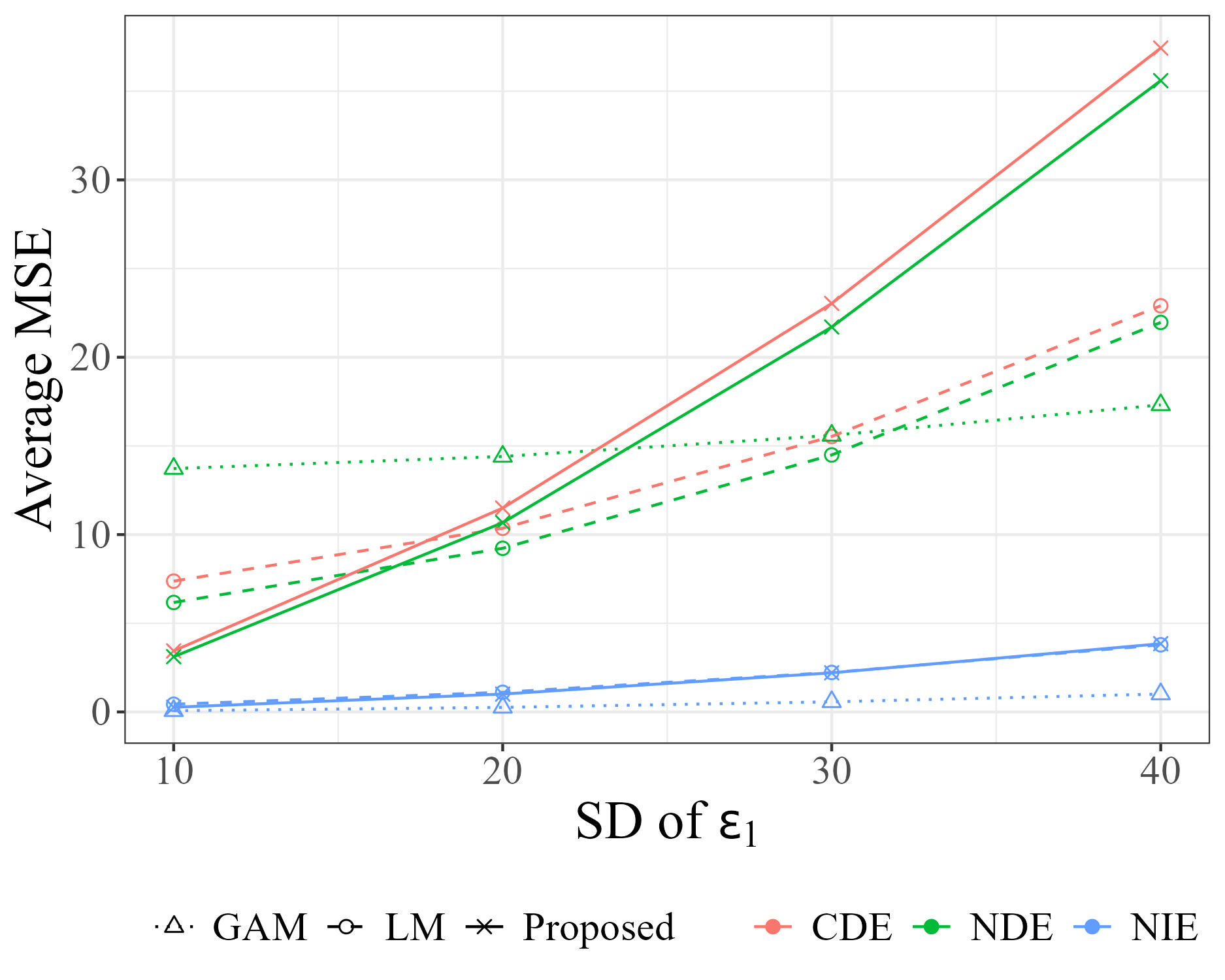}}\par
\subfloat[Average $|$Relative Bias$|$ under pattern 3]{\label{fig:3:3.c}\includegraphics[width=.45\linewidth]{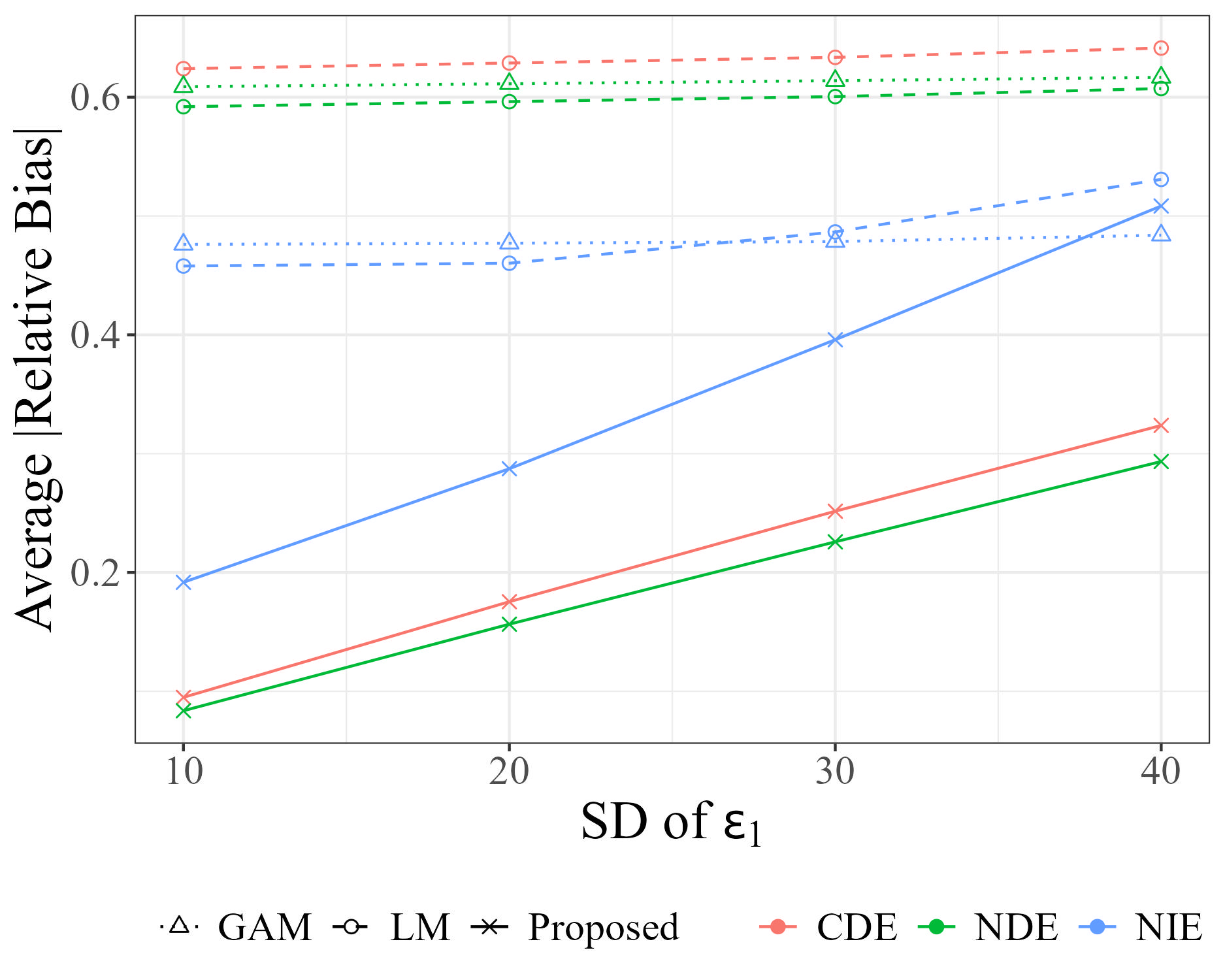}}\hfill
\subfloat[Average MSE under pattern 3]{\label{fig:3:3.d}\includegraphics[width=.45\linewidth]{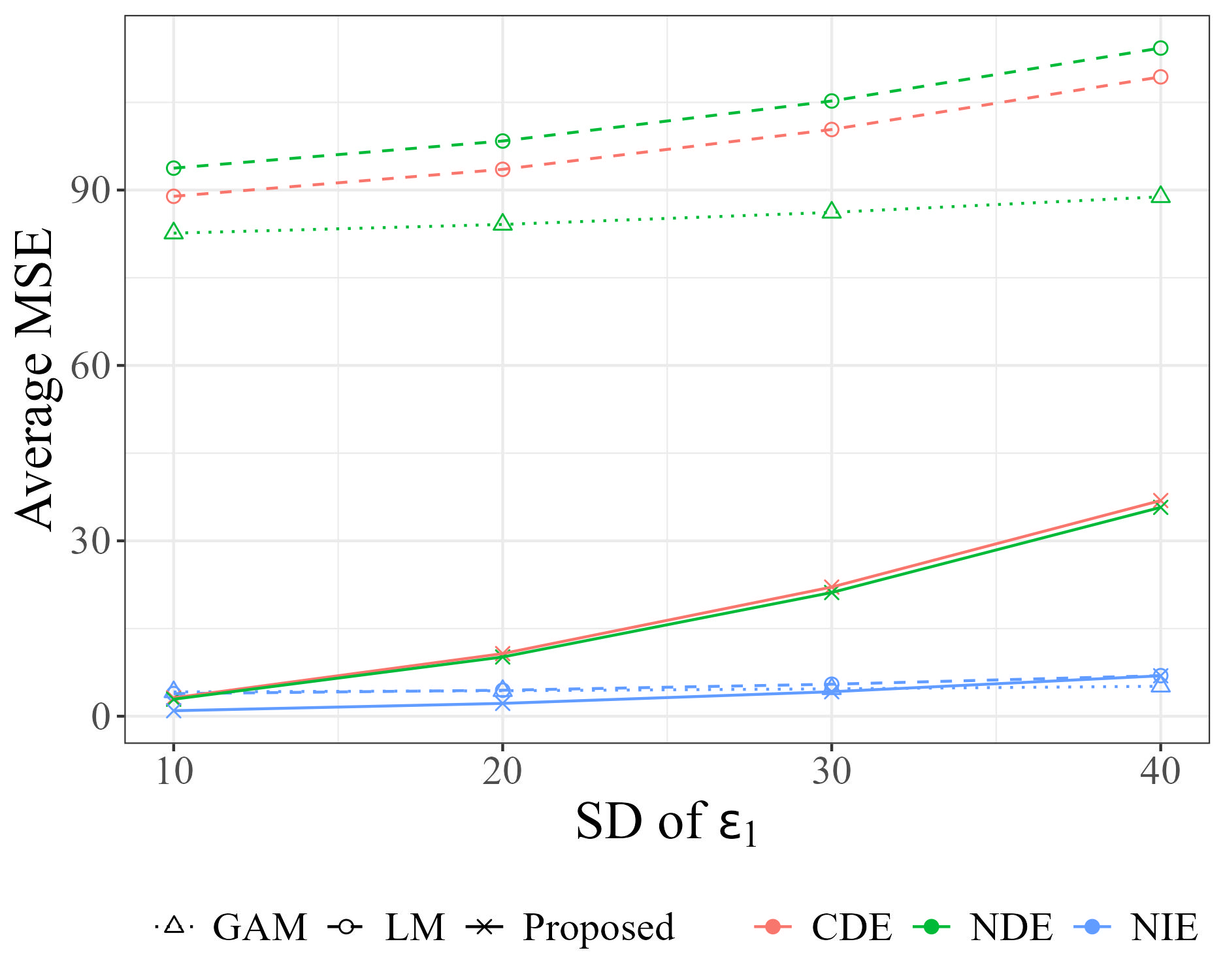}}\par
\caption{Simulation results of average absolute relative bias and average MSE for each pattern}
\label{fig:3:1-3.c-d}
\end{figure}

\subsection*{S.6. Plots of hormone vs. birth weight under linear pattern and simulation results for linear pattern}
\begin{figure}[H]
\centering
\subfloat[Linear Pattern]{\label{fig:3:4.a}\includegraphics[width=.45\linewidth]{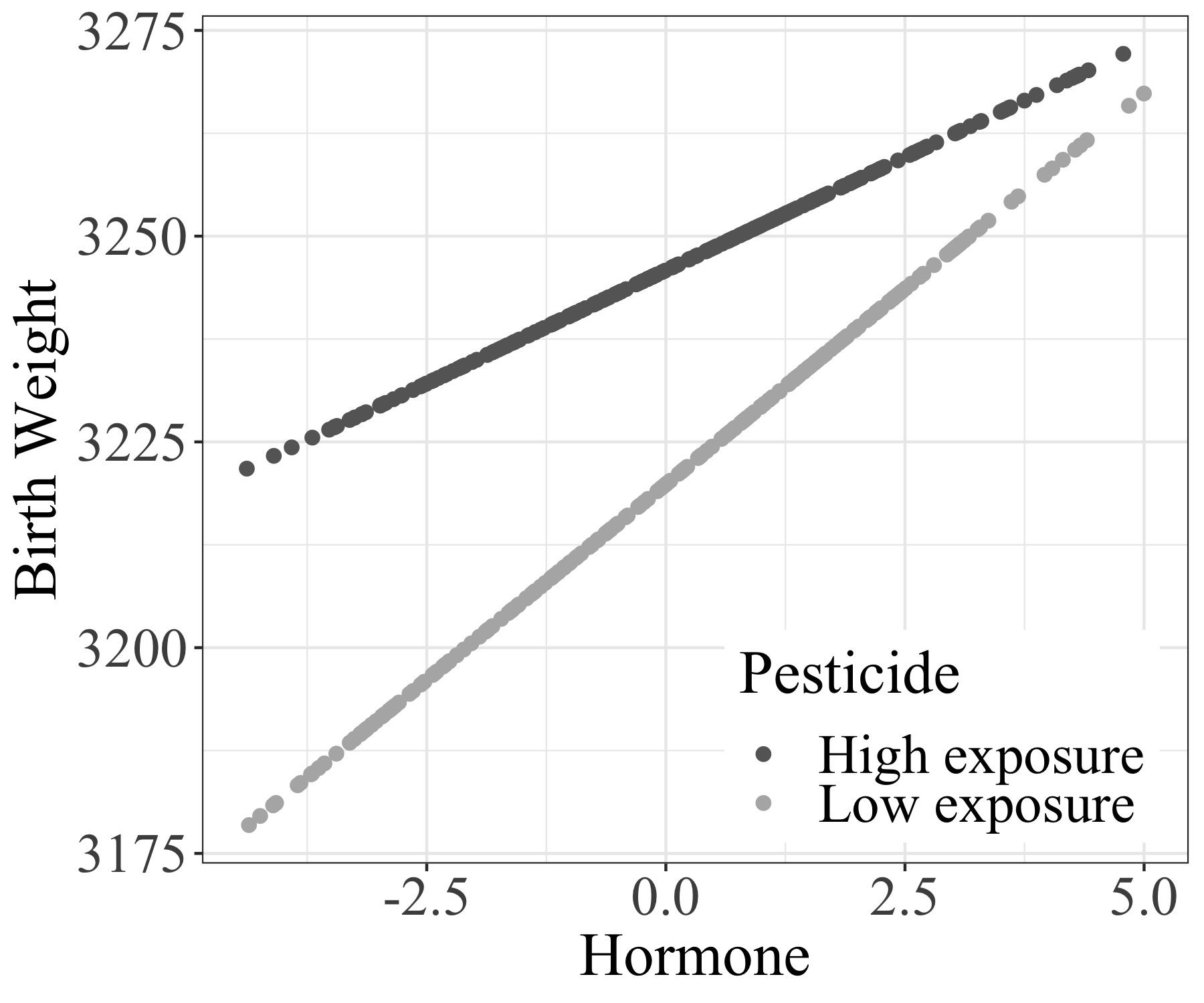}}\hfill
\subfloat[Coverage Probability]{\label{fig:3:4.b}\includegraphics[width=.45\linewidth]{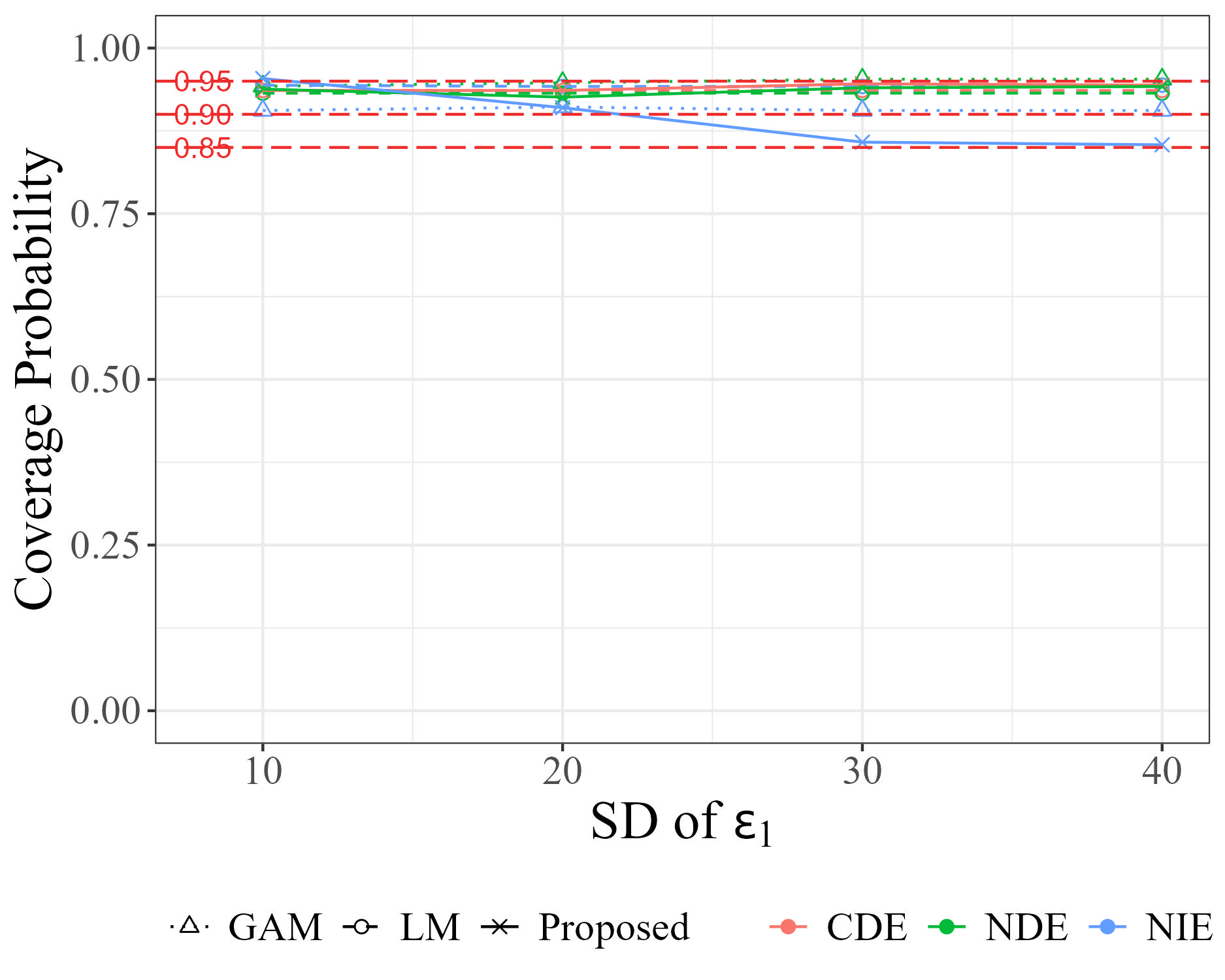}}\par
\subfloat[Average $|$Relative Bias$|$]{\label{fig:3:4.c}\includegraphics[width=.45\linewidth]{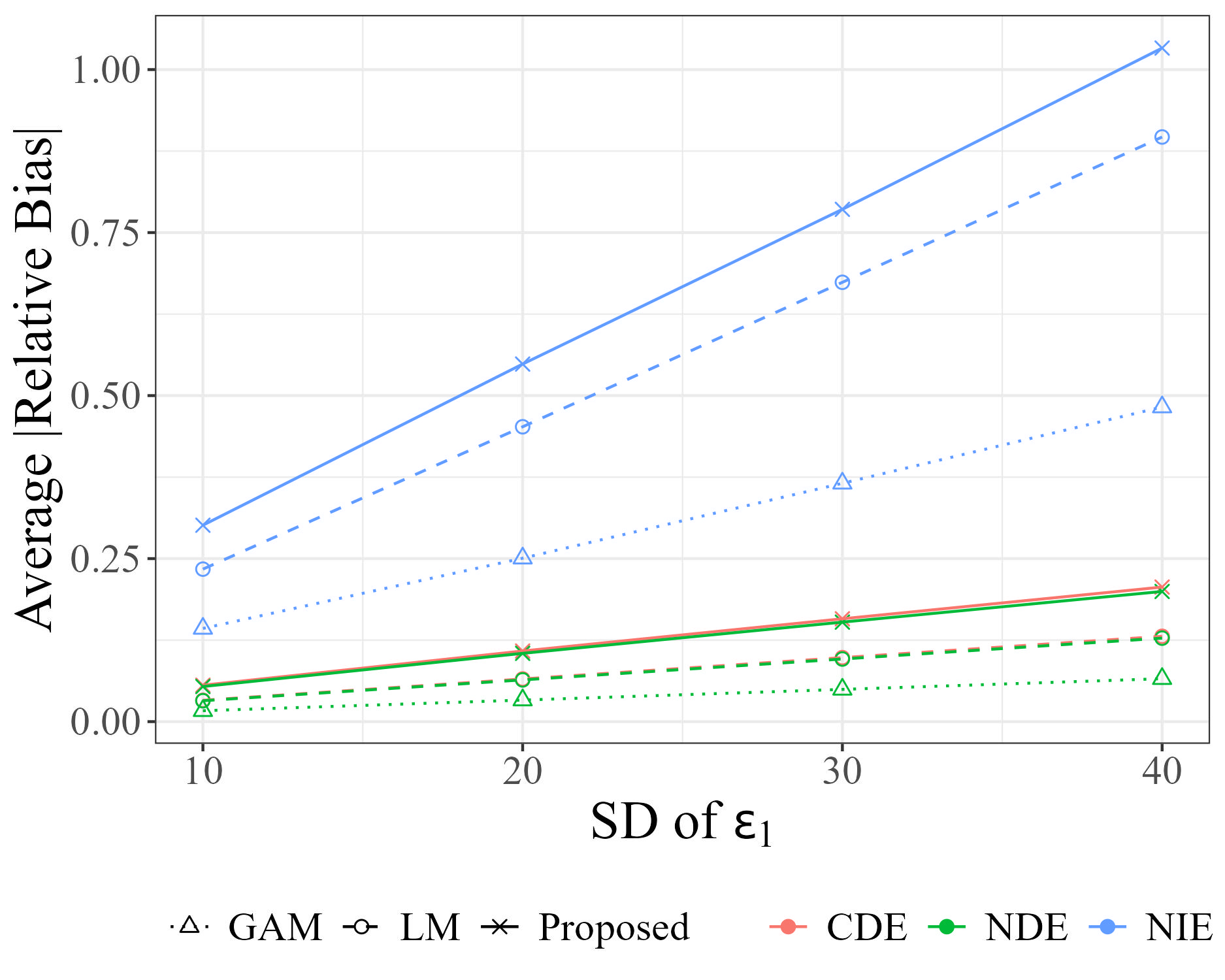}}\hfill
\subfloat[Average MSE]{\label{fig:3:4.d}\includegraphics[width=.45\linewidth]{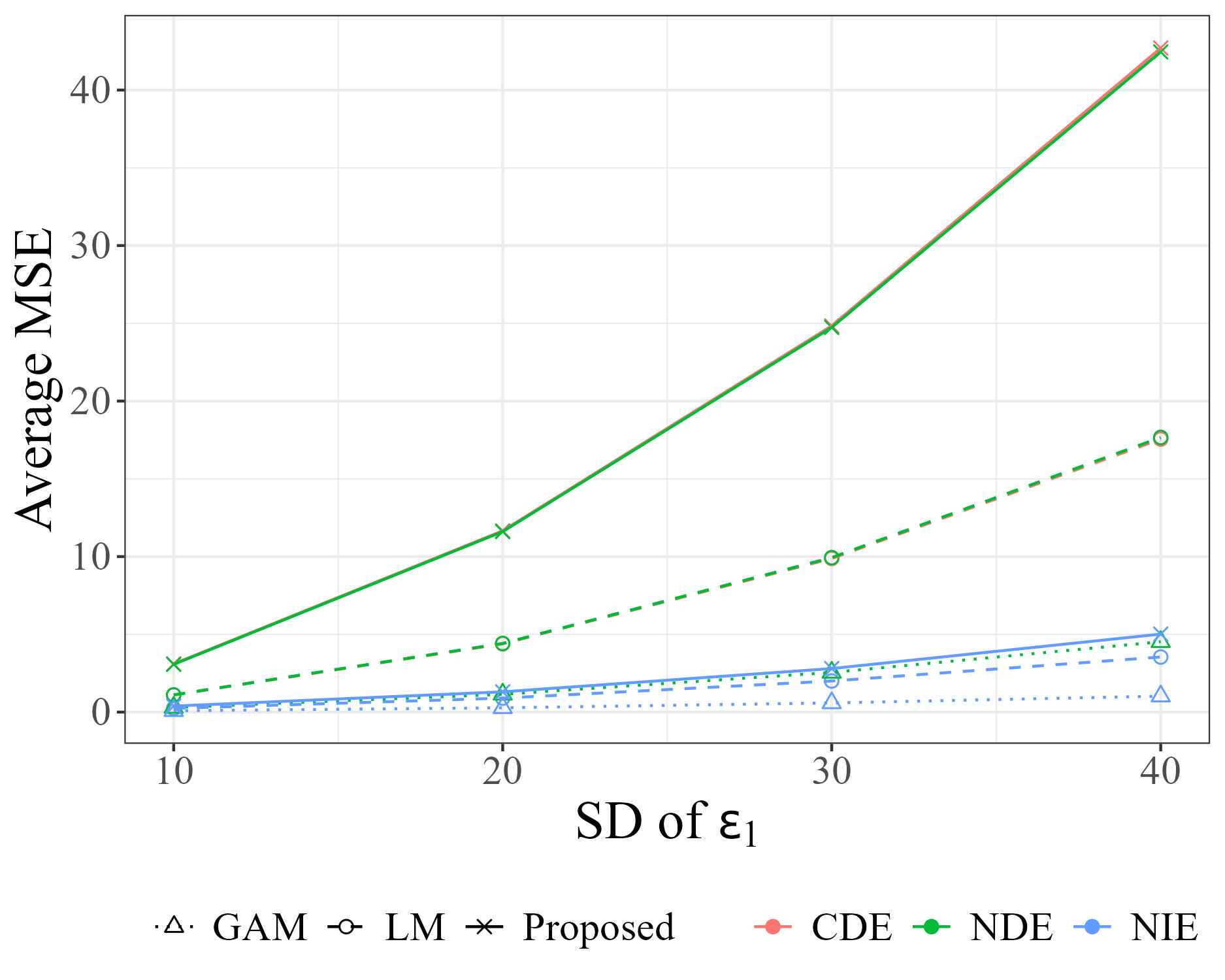}}\par
\caption{Plots of hormone vs. birth weight under linear pattern, and simulation results of coverage probability, average absolute relative bias and average MSE for linear pattern}
\label{fig:3:4}
\end{figure}

\subsection*{S.7. Tables of Simulation Results}
\begin{table}
\begin{center}
\caption{Simulation results of coverage probability, average absolute relative bias and average MSE for pattern 1 (true CDE: $\sim$ 44.62, true NDE: 45.82, true NIE: 1.10)}
\label{tab:3:2}
\begin{tabular}{ c c c c c c c c c c } 
\hline
\multicolumn{10}{c}{Semi-parametric shape-restricted regression spline} \\
\hline
Variance of $\epsilon_1$ & \multicolumn{3}{c}{Coverage Probability} & \multicolumn{3}{c}{Average $|$Relative Bias$|$} & \multicolumn{3}{c}{Average MSE} \\
\cline{2-10}
 & CDE & NDE & NIE & CDE & NDE & NIE & CDE & NDE & NIE \\
\hline
$10^2$ & 0.922 & 0.948 & 0.958 & 0.032 & 0.030 & 0.374 & 3.223 & 2.879 & 0.281 \\ 
\hline
$20^2$ & 0.932 & 0.948 & 0.950 & 0.059 & 0.056 & 0.661 & 10.808 & 10.030 & 0.902 \\ 
\hline
$30^2$ & 0.954 & 0.952 & 0.958 & 0.084 & 0.080 & 0.945 & 21.982 & 20.739 & 1.852 \\ 
\hline
$40^2$ & 0.952 & 0.952 & 0.960 & 0.108 & 0.103 & 1.232 & 36.465 & 34.837 & 3.129 \\ 
\hline
\hline
\multicolumn{10}{c}{Linear regression} \\
\hline
Variance of $\epsilon_1$ & \multicolumn{3}{c}{Coverage Probability} & \multicolumn{3}{c}{Average $|$Relative Bias$|$} & \multicolumn{3}{c}{Average MSE} \\
\cline{2-10}
 & CDE & NDE & NIE & CDE & NDE & NIE & CDE & NDE & NIE \\
\hline
$10^2$ & 0.000 & 0.000 & 0.922 & 0.199 & 0.204 & 0.402 & 79.924 & 88.628 & 0.312 \\ 
\hline
$20^2$ & 0.018 & 0.008 & 0.928 & 0.197 & 0.203 & 0.696 & 81.953 & 90.648 & 0.959 \\ 
\hline
$30^2$ & 0.198 & 0.158 & 0.936 & 0.196 & 0.201 & 1.016 & 86.176 & 94.878 & 2.048 \\ 
\hline
$40^2$ & 0.444 & 0.400 & 0.940 & 0.195 & 0.200 & 1.343 & 92.594 & 101.316 & 3.580 \\ 
\hline
\hline
\multicolumn{10}{c}{Generalized additive model} \\
\hline
Variance of $\epsilon_1$ & \multicolumn{3}{c}{Coverage Probability} & \multicolumn{3}{c}{Average $|$Relative Bias$|$} & \multicolumn{3}{c}{Average MSE} \\
\cline{2-10}
 & CDE & NDE & NIE & CDE & NDE & NIE & CDE & NDE & NIE \\
\hline
$10^2$ & - & 0.000 & 0.895 & - & 0.188 & 0.210 & - & 70.406 & 0.072 \\ 
\hline
$20^2$ & - & 0.000 & 0.901 & - & 0.187 & 0.394 & - & 70.662 & 0.257 \\ 
\hline
$30^2$ & - & 0.005 & 0.901 & - & 0.186 & 0.583 & - & 71.381 & 0.565 \\ 
\hline
$40^2$ & - & 0.016 & 0.911 & - & 0.186 & 0.775 & - & 72.631 & 0.996 \\ 
\hline
\end{tabular}
\end{center}
\end{table}

\begin{table}
\begin{center}
\caption{Simulation results of coverage probability, average absolute relative bias and average MSE for pattern 2 (true CDE: $\sim$ 19.85, true NDE: 19.17, true NIE: -0.17)}
\label{tab:3:3}
\begin{tabular}{ c c c c c c c c c c } 
\hline
\multicolumn{10}{c}{Semi-parametric shape-restricted regression spline} \\
\hline
Variance of $\epsilon_1$ & \multicolumn{3}{c}{Coverage Probability} & \multicolumn{3}{c}{Average $|$Relative Bias$|$} & \multicolumn{3}{c}{Average MSE} \\
\cline{2-10}
 & CDE & NDE & NIE & CDE & NDE & NIE & CDE & NDE & NIE \\
\hline
$10^2$ & 0.950 & 0.974 & 0.972 & 0.076 & 0.074 & 2.405 & 3.442 & 3.114 & 0.264 \\ 
\hline
$20^2$ & 0.948 & 0.966 & 0.968 & 0.138 & 0.137 & 4.704 & 11.496 & 10.673 & 1.008 \\ 
\hline
$30^2$ & 0.944 & 0.956 & 0.962 & 0.195 & 0.195 & 6.954 & 23.035 & 21.703 & 2.205 \\ 
\hline
$40^2$ & 0.946 & 0.962 & 0.962 & 0.247 & 0.249 & 9.200 & 37.434 & 35.597 & 3.847 \\ 
\hline
\hline
\multicolumn{10}{c}{Linear regression} \\
\hline
Variance of $\epsilon_1$ & \multicolumn{3}{c}{Coverage Probability} & \multicolumn{3}{c}{Average $|$Relative Bias$|$} & \multicolumn{3}{c}{Average MSE} \\
\cline{2-10}
 & CDE & NDE & NIE & CDE & NDE & NIE & CDE & NDE & NIE \\
\hline
$10^2$ & 0.560 & 0.638 & 0.928 & 0.124 & 0.116 & 3.157 & 7.376 & 6.172 & 0.434 \\ 
\hline
$20^2$ & 0.796 & 0.836 & 0.940 & 0.134 & 0.130 & 4.953 & 10.357 & 9.227 & 1.109 \\ 
\hline
$30^2$ & 0.888 & 0.892 & 0.938 & 0.159 & 0.159 & 7.000 & 15.533 & 14.491 & 2.226 \\ 
\hline
$40^2$ & 0.908 & 0.928 & 0.938 & 0.192 & 0.195 & 9.135 & 22.902 & 21.964 & 3.786 \\ 
\hline
\hline
\multicolumn{10}{c}{Generalized additive model} \\
\hline
Variance of $\epsilon_1$ & \multicolumn{3}{c}{Coverage Probability} & \multicolumn{3}{c}{Average $|$Relative Bias$|$} & \multicolumn{3}{c}{Average MSE} \\
\cline{2-10}
 & CDE & NDE & NIE & CDE & NDE & NIE & CDE & NDE & NIE \\
\hline
$10^2$ & - & 0.000 & 0.895 & - & 0.186 & 0.619 & - & 13.722 & 0.068 \\ 
\hline
$20^2$ & - & 0.037 & 0.901 & - & 0.185 & 1.195 & - & 14.407 & 0.257 \\ 
\hline
$30^2$ & - & 0.356 & 0.901 & - & 0.185 & 1.771 & - & 15.593 & 0.571 \\ 
\hline
$40^2$ & - & 0.550 & 0.901 & - & 0.187 & 2.351 & - & 17.313 & 1.010 \\ 
\hline
\end{tabular}
\end{center}
\end{table}

\begin{table}
\begin{center}
\caption{Simulation results of coverage probability, average absolute relative bias and average MSE for pattern 3 (true CDE: $\sim$ -15.00, true NDE: -16.24, true NIE: 4.08)}
\label{tab:3:4}
\begin{tabular}{ c c c c c c c c c c } 
\hline
\multicolumn{10}{c}{Semi-parametric shape-restricted regression spline} \\
\hline
Variance of $\epsilon_1$ & \multicolumn{3}{c}{Coverage Probability} & \multicolumn{3}{c}{Average $|$Relative Bias$|$} & \multicolumn{3}{c}{Average MSE} \\
\cline{2-10}
 & CDE & NDE & NIE & CDE & NDE & NIE & CDE & NDE & NIE \\
\hline
$10^2$ & 0.920 & 0.944 & 0.894 & 0.095 & 0.084 & 0.192 & 3.145 & 2.910 & 0.940 \\ 
\hline
$20^2$ & 0.934 & 0.946 & 0.924 & 0.175 & 0.156 & 0.287 & 10.708 & 10.134 & 2.193 \\ 
\hline
$30^2$ & 0.940 & 0.950 & 0.934 & 0.252 & 0.226 & 0.396 & 22.092 & 21.157 & 4.190 \\ 
\hline
$40^2$ & 0.942 & 0.954 & 0.936 & 0.324 & 0.293 & 0.508 & 36.892 & 35.725 & 6.915 \\ 
\hline
\hline
\multicolumn{10}{c}{Linear regression} \\
\hline
Variance of $\epsilon_1$ & \multicolumn{3}{c}{Coverage Probability} & \multicolumn{3}{c}{Average $|$Relative Bias$|$} & \multicolumn{3}{c}{Average MSE} \\
\cline{2-10}
 & CDE & NDE & NIE & CDE & NDE & NIE & CDE & NDE & NIE \\
\hline
$10^2$ & 0.000 & 0.000 & 0.130 & 0.624 & 0.592 & 0.458 & 88.930 & 93.747 & 3.865 \\ 
\hline
$20^2$ & 0.010 & 0.010 & 0.532 & 0.629 & 0.596 & 0.460 & 93.545 & 98.385 & 4.448 \\ 
\hline
$30^2$ & 0.148 & 0.134 & 0.736 & 0.634 & 0.601 & 0.487 & 100.355 & 105.232 & 5.473 \\ 
\hline
$40^2$ & 0.352 & 0.336 & 0.820 & 0.641 & 0.607 & 0.531 & 109.360 & 114.288 & 6.940 \\ 
\hline
\hline
\multicolumn{10}{c}{Generalized additive model} \\
\hline
Variance of $\epsilon_1$ & \multicolumn{3}{c}{Coverage Probability} & \multicolumn{3}{c}{Average $|$Relative Bias$|$} & \multicolumn{3}{c}{Average MSE} \\
\cline{2-10}
 & CDE & NDE & NIE & CDE & NDE & NIE & CDE & NDE & NIE \\
\hline
$10^2$ & - & 0.000 & 0.000 & - & 0.609 & 0.476 & - & 82.627 & 4.150 \\ 
\hline
$20^2$ & - & 0.000 & 0.068 & - & 0.611 & 0.477 & - & 84.116 & 4.360 \\ 
\hline
$30^2$ & - & 0.000 & 0.288 & - & 0.614 & 0.479 & - & 86.181 & 4.684 \\ 
\hline
$40^2$ & - & 0.010 & 0.503 & - & 0.617 & 0.484 & - & 88.846 & 5.125 \\ 
\hline
\end{tabular}
\end{center}
\end{table}

\begin{table}
\begin{center}
\caption{Simulation results of coverage probability, average absolute relative bias and average MSE for linear pattern (true CDE: $\sim$ 25.44, true NDE: 26.07, true NIE: 1.65)}
\label{tab:3:5}
\begin{tabular}{ c c c c c c c c c c } 
\hline
\multicolumn{10}{c}{Semi-parametric shape-restricted regression spline} \\
\hline
Variance of $\epsilon_1$ & \multicolumn{3}{c}{Coverage Probability} & \multicolumn{3}{c}{Average $|$Relative Bias$|$} & \multicolumn{3}{c}{Average MSE} \\
\cline{2-10}
 & CDE & NDE & NIE & CDE & NDE & NIE & CDE & NDE & NIE \\
\hline
$10^2$ & 0.936 & 0.938 & 0.954 & 0.056 & 0.054 & 0.301 & 1.421 & 1.404 & 0.497 \\ 
\hline
$20^2$ & 0.936 & 0.926 & 0.910 & 0.108 & 0.105 & 0.548 & 2.755 & 2.730 & 0.905 \\ 
\hline
$30^2$ & 0.946 & 0.940 & 0.858 & 0.158 & 0.153 & 0.786 & 4.012 & 3.978 & 1.296 \\ 
\hline
$40^2$ & 0.944 & 0.942 & 0.854 & 0.206 & 0.200 & 1.033 & 5.249 & 5.204 & 1.704 \\ 
\hline
\hline
\multicolumn{10}{c}{Linear regression} \\
\hline
Variance of $\epsilon_1$ & \multicolumn{3}{c}{Coverage Probability} & \multicolumn{3}{c}{Average $|$Relative Bias$|$} & \multicolumn{3}{c}{Average MSE} \\
\cline{2-10}
 & CDE & NDE & NIE & CDE & NDE & NIE & CDE & NDE & NIE \\
\hline
$10^2$ & 0.936 & 0.932 & 0.944 & 0.033 & 0.032 & 0.234 & 1.097 & 1.107 & 0.239 \\ 
\hline
$20^2$ & 0.936 & 0.932 & 0.942 & 0.065 & 0.064 & 0.452 & 4.389 & 4.417 & 0.898 \\ 
\hline
$30^2$ & 0.936 & 0.932 & 0.942 & 0.098 & 0.096 & 0.674 & 9.875 & 9.936 & 1.999 \\ 
\hline
$40^2$ & 0.936 & 0.932 & 0.944 & 0.131 & 0.128 & 0.897 & 17.556 & 17.664 & 3.543 \\ 
\hline
\hline
\multicolumn{10}{c}{Generalized additive model} \\
\hline
Variance of $\epsilon_1$ & \multicolumn{3}{c}{Coverage Probability} & \multicolumn{3}{c}{Average $|$Relative Bias$|$} & \multicolumn{3}{c}{Average MSE} \\
\cline{2-10}
 & CDE & NDE & NIE & CDE & NDE & NIE & CDE & NDE & NIE \\
\hline
$10^2$ & - & 0.942 & 0.906 & - & 0.017 & 0.143 & - & 0.295 & 0.088 \\ 
\hline
$20^2$ & - & 0.948 & 0.911 & - & 0.033 & 0.251 & - & 1.143 & 0.276 \\ 
\hline
$30^2$ & - & 0.953 & 0.906 & - & 0.049 & 0.366 & - & 2.553 & 0.589 \\ 
\hline
$40^2$ & - & 0.953 & 0.906 & - & 0.066 & 0.482 & - & 4.523 & 1.026 \\ 
\hline
\end{tabular}
\end{center}
\end{table}